\documentclass[manuscript,screen]{acmart}
\setcopyright{none}

\usepackage{forest}
\useforestlibrary{edges}
\usepackage{enumitem}
\usepackage{cleveref}
\usepackage{fontawesome5} % for the lightbulb icon
\usepackage{multirow}
\usepackage{colortbl}

\usepackage{tikz}
\usepackage{booktabs}

\usepackage{framed}
\usepackage{xcolor}
\definecolor{rqframe}{HTML}{19365C}  % ~ blue!60!black
\definecolor{rqback}{HTML}{EDF1FB}   % ~ blue!5
\definecolor{shadecolor}{named}{rqback}
\colorlet{defframe}{orange!50!black}  % identical colours to your tcolorbox
\colorlet{defback}{yellow!10}

\newcounter{defn}
\crefname{defn}{Definition}{Definitions} 

\newlength{\taxRootW}
\newlength{\taxLevelOneW}
\newlength{\taxLevelTwoW}
\newlength{\taxRefsW}

\forestset{
  box/.style={
    draw=teal!55!black,
    rounded corners=1pt,
    line width=0.6pt,
    fill=white,
    inner sep=3pt,
    align=center,
    font=\footnotesize,
    text width=\taxLevelTwoW, % default width for "normal" boxes
  },
  rootbox/.style={
    box,
    fill=teal!55!black,
    text=white,
    text width=\taxRootW,
  },
  levelonebox/.style={
    box,
    fill=teal!20!white,
    text width=\taxLevelOneW,
  },
  leveltwobox/.style={
    box,
    fill=teal!10!white,
    text width=\taxLevelTwoW,
  },
  refs/.style={
    draw=teal!55!black,
    rounded corners=3pt,
    line width=0.6pt,
    fill=white,
    inner sep=1.5pt,
    align=left,
    font=\footnotesize,
    text width=\taxRefsW,
  },
  refswide/.style={
    refs,
    text width=\dimexpr\taxLevelTwoW+\taxRefsW+5mm\relax,
  },
  taxonomy tree/.style={
    for tree={
      grow'=east,
      parent anchor=east,
      child anchor=west,
      anchor=west,
      l sep=2mm,
      s sep=1mm,
      forked edges,
      fork sep=1mm,
    }
  },
  edge/.style={
    draw=black!55,
    line width=0.45pt,
  },
}

\colorlet{defframe}{orange!50!black}
\colorlet{defback}{yellow!10}
\colorlet{rqframe}{blue!60!black}
\colorlet{rqback}{blue!5}

\makeatletter
\newcommand{\@boxtitle}[2]{%
  \vspace{-\FrameSep}%
  \noindent\hspace*{-\FrameSep}%
  {\setlength{\fboxsep}{\FrameSep}%
   \colorbox{#1}{\makebox[\linewidth][l]{\color{white}\bfseries #2}}}%
  \par\vspace{\dimexpr\FrameSep-1pt}\noindent\ignorespaces}

\newenvironment{definition}[2]
  {\refstepcounter{defn}\label{#2}\par\smallskip
   \colorlet{shadecolor}{defback}\begin{shaded}%
   \@boxtitle{defframe}{\faBook\ Definition~\thedefn: #1}}
  {\end{shaded}\smallskip}

\newenvironment{rqfindings}[1]
  {\par\smallskip
   \colorlet{shadecolor}{rqback}\begin{shaded}%
   \@boxtitle{rqframe}{\faListUl\ Key Findings RQ#1}}
  {\end{shaded}\smallskip}
\makeatother

\begin{document}

\title{Multi-Agent Debate Strategies: Survey, Taxonomy, and Challenges}

\author{Quim Motger}
\affiliation{%
  \institution{Dpt. of Service and Information System Engineering, Universitat Polit\`ecnica de Catalunya}
  \city{Barcelona}
  \country{Spain}
}
\email{joaquim.motger@upc.edu}

\author{Marc Oriol}
\affiliation{%
  \institution{Dpt. of Service and Information System Engineering, Universitat Polit\`ecnica de Catalunya}
  \city{Barcelona}
  \country{Spain}
}
\email{marc.oriol@upc.edu}

\author{Jordi Marco}
\affiliation{%
  \institution{Dpt. of Computer Science, Universitat Polit\`ecnica de Catalunya}
  \city{Barcelona}
  \country{Spain}
}
\email{jordi.marco@upc.edu}

\author{Xavier Franch}
\affiliation{%
  \institution{Dpt. of Service and Information System Engineering, Universitat Polit\`ecnica de Catalunya}
  \city{Barcelona}
  \country{Spain}
}
\email{xavier.franch@upc.edu}

\renewcommand{\shortauthors}{Motger et al.}
%%
%% Article type: Research, Review, Discussion, Invited or position
\acmArticleType{Review}
%%
%% Links to code and data
% \acmCodeLink{https://github.com/borisveytsman/acmart}
% \acmDataLink{htps://zenodo.org/link}
% %%
% %% Authors' contribution
% \acmContributions{TO-DO}



\begin{abstract}
% CONTEXT
Multi-Agent Debate (MAD) is a promising paradigm for improving the accuracy and robustness of Large Language Model (LLM)-based agentic systems. It enables multiple agents to exchange arguments, critique each other's outputs, and iteratively converge towards a solution. %, MAD offers a data-efficient alternative to fine-tuning and retrieval-augmented approaches. 
% PROBLEM
However, research remains fragmented, with inconsistent terminology and no rigorous synthesis of MAD design dimensions. 
% METHOD
We present a systematic literature review characterizing 141 primary studies on MAD. We derive a three-dimensional taxonomy covering debate participants, the interaction mechanisms structuring the exchange, and the agreement protocols governing debate resolution, supported by formal notations to render MAD configurations. 
% RESULTS
Our analysis reveals that the field has implicitly converged on a narrow design pattern --- static, fully connected topologies, verbatim exchange, short-term memory and voting resolution strategies --- adopted by convention rather than systematic comparison, while promising alternatives remain marginal. Because any MAD setting reflects roughly a dozen interacting design decisions, cross-study comparison is unreliable when these are left implicit. 
% CONCLUSIONS
We position the taxonomy as a descriptive map of the research landscape, a framework for controlled benchmarking, and potentially as a schema for machine-readable MAD specifications. As future work, we propose formalizing it into an executable specification, enabling cost-aware benchmarking and automated tuning of debate configurations.
\end{abstract}


\maketitle

\section{Introduction}
\label{sec:introduction}

The adoption of Large Language Models (LLMs) within agentic systems has rapidly evolved from a conceptual research direction into a concrete innovation pathway for industrial software-intensive systems~\cite{Xi2025}. 
In this paradigm, an \emph{agent} is commonly defined as an LLM-based autonomous system that leverages reasoning, planning, memory, and tool-use capabilities to perceive contextual inputs and generate goal-directed actions in dynamic environments~\cite{Wang2024}. 
LLM-based agents iteratively observe, reason, and act through structured interaction loops~\cite{Yao2023ReAct}, enabling explicit interleaving of reasoning traces and contextual feedback. 
When multiple autonomous LLM-based agents interact within a shared environment, the resulting configuration constitutes an LLM-driven \emph{multi-agent system}, characterized by distributed cognition, role specialization, communication via natural language, and collaborative problem solving~\cite{park2023generative,Hong2024MetaGPT}. 
Agents pursue shared or complementary \emph{goals}, operationalized as task objectives decomposed into coordinated subtasks, often supported by reflection mechanisms, planning modules, memory components, and structured inter-agent dialogue protocols~\cite{Yao2023ReAct,Xi2025}. 

Empirical research in LLM-driven multi-agent systems has mainly examined how architectural and adaptation strategies affect task performance and efficiency, largely by strengthening individual agents or system-level reliability through established LLM adaptation paradigms.
At inference time, prompting-based methods elicit latent capabilities without modifying parameters, including few-shot in-context learning~\cite{li-etal-2023-shot}, chain-of-thought reasoning~\cite{Wei2022CoT}, self-consistency decoding~\cite{Wang2023SelfConsistency}, and tree-of-thought search~\cite{Yao2023TreeOfThought}. 
Other paradigms improve goal adherence and reliability through alignment, such as reinforcement learning from human feedback and preference optimization~\cite{Ouyang2023RLHF,Rafailov2023DPO}; parameter adaptation, including full or parameter-efficient fine-tuning~\cite{Dettmers2023QLoRA} and continued domain-adaptive pre-training~\cite{cheng2024adaptinglargelanguagemodels}; or retrieval-augmented generation, which injects external knowledge to improve grounding and factual consistency~\cite{Fan2024}. 
In multi-agent settings, these techniques are often embedded into role-specialized collaborative frameworks or agent societies~\cite{Hong2024MetaGPT}. 
However, except for some prompting strategies, these paradigms typically introduce additional resource requirements, such as domain-specific corpora for continued pre-training~\cite{Gururangan2020}, curated knowledge bases for retrieval-augmented generation~\cite{Zhao2026}, or human preference annotations for alignment~\cite{Retzlaff2024}.

In this context, improving the accuracy and robustness of LLM-based agents without additional domain-specific data remains a central challenge.
A promising direction draws inspiration from human collaboration. 
People often improve decisions by exchanging viewpoints, debating inconsistencies, and converging towards consensus. 
Similarly, multiple LLM-based agents can interact, critique each other’s outputs, and iteratively refine their reasoning. 
This paradigm, known as Multi-Agent Debate (MAD)~\cite{12, 3}, has gained increasing interest with the rise of advanced LLMs.% MAD has been successfully applied to tasks such as solving complex mathematical problems~\cite{p03-li-2024} and improving performance on general knowledge benchmarks~\cite{p17-wang-2025}. 

Despite this growing body of work, the field remains fragmented, with limited consolidation of terminology, design dimensions, and architectural patterns across application domains. 
Most literature reviews treat debate primarily as a descriptor of specific interaction mechanisms~\cite{Li2024a, Tran2025, Wang2026}, lacking a deeper and more systematic characterization. 
Several studies qualitatively report approaches that implement debate strategies~\cite{Shen2023, Li2024, Tran2025, Luo2025, Guan2026}, yet they do not provide a structured analysis or a comprehensive elicitation of core debate properties. 
Smit et al.~\cite{6} offered an early overview of MAD research, identifying eight studies and extracting six design features, such as the number of debate rounds and the inclusion of specialized roles like judge or summarizer. 
Wang et al. proposed a flat categorization of MAD strategies based on three communication patterns~\cite{Wang2026}. 
Tillmann et al.~\cite{Tillmann2025} also reviewed MAD approaches, although their review lacks a clear systematic protocol, relies exclusively on an AI-assisted tool, and does not report the exact number of included studies. 
\textbf{To the best of our knowledge, no study provides a large-scale, methodologically rigorous, and cross-domain synthesis that systematically consolidates MAD design dimensions and architectural patterns.}

In our previous study, we conducted an initial systematic mapping to identify and analyse MAD strategies across multiple domains, focusing on their impact and value within the requirements engineering (RE) field~\cite{Oriol2025}. 
We derived a tentative taxonomy and a consolidated vocabulary based on 25 primary studies. 
In this study, and following the ACM SIGSOFT Empirical Standards for systematic reviews~\cite{SIGSOFT}, we build on that foundation to broaden the scope of the review through backward and forward snowballing and an improved data extraction protocol. 
We refine, extend, and validate the taxonomy while expanding the review both in breadth and depth, enlarging the set of surveyed studies across domains and providing a substantially more detailed and systematic analysis of MAD design dimensions and mechanisms. 

Specifically, this study addresses the following research questions (RQs): 

\textbf{RQ1.} %What is the research landscape of MAD in terms of application domains, addressed tasks, research contexts, and terminology used in the literature? 
What is the research \emph{landscape} of MAD in terms of application domains and addressed tasks?

\textbf{RQ2.} How are \emph{participants} designed in MAD approaches, and what mechanisms are employed to define their roles, capabilities, and individual decision-making behaviours?

\textbf{RQ3.} How are \emph{interaction} processes structured in MAD approaches, and what mechanisms are employed to coordinate communication, information exchange, and influence among agents?

\textbf{RQ4.} How is \emph{agreement} achieved in MAD approaches, and what resolution mechanisms are employed to determine final outcomes?

Through the assessment of these RQs, the study delivers the following contributions\footnote{All research materials are publicly available in a replication package. See \nameref{sec:das} at the end of this manuscript.}:

\begin{enumerate}[label=\textbf{C\arabic*.}]
    \item A systematic literature review of MAD, characterizing 141 primary studies and providing a structured synthesis of the research landscape.
    \item A three-dimensional taxonomy of MAD design, formalizing practical design and development dimensions across: (i) \emph{participants} involved in the debate, (ii) \emph{interaction} mechanisms structuring the exchange, and (iii) \emph{agreement} protocols governing debate resolution.
    \item A consolidated analysis of open challenges and emerging research opportunities, outlining directions for advancing MAD methodologies and applications.
\end{enumerate}

The remainder of this paper is structured as follows. 
Section \ref{sec:search-methodology} describes the methodology of the systematic literature review and presents the search results. 
The results of the data extraction and analysis of the surveyed studies are presented across several sections. 
Specifically, the research landscape is presented in Section \ref{sec:landscape} (RQ1), followed by the taxonomies derived for each MAD design dimension: participants (Section \ref{sec:participants}, RQ2), interaction (Section \ref{sec:interaction}, RQ3), and agreement (Section \ref{sec:type-of-agreement}, RQ4). 
Section~\ref{sec:discussion} synthesizes the main implications of the taxonomy, highlighting dominant design patterns, underexplored alternatives, and methodological challenges for comparing and benchmarking MAD approaches.
Section \ref{sec:threats-to-validity} describes threats to validity of the study. 
Finally, the main conclusions and future work are presented in Section \ref{sec:conclusions}.

\section{Method}
\label{sec:search-methodology}

This study follows the systematic literature review (SLR) guidelines proposed by Kitchenham and Charters~\cite{Kitchenham2007}, structured around an explicit search and study-selection protocol, coupled with a snowballing procedure as formalized by Wohlin~\cite{Wohlin2014} to complement the database search with backward and forward snowballing. The overall process comprises five stages: (i) definition and scoping of MAD, (ii) definition of a seed search string, (iii) independent screening of candidate papers against explicit inclusion/exclusion criteria, (iv)  backward and forward snowballing, and (v) structured data extraction and inductive coding to derive the taxonomy.

\subsection{Definition and Scope}
\label{sec:mad-definition}
Given the diversity of interpretations and terminology observed across the literature, it is essential to establish a clear and consistent definition of MAD that delineates what falls within the scope of this study. 

We begin by examining the foundational concept of debate. Debate is defined in the Cambridge Dictionary\footnote{https://dictionary.cambridge.org} as a ``\emph{serious discussion of a subject in which many people take part}'', where discussion is ``\emph{the activity in which people talk about something and tell each other their ideas or opinions}''. 
While traditional debates involve human participants exchanging arguments and counterarguments, the same concept can be applied to Multi-Agent Systems (MAS). MAD can be intuitively understood as a particular interaction paradigm within a MAS, in which multiple agents engage in a structured exchange of arguments or opinions.
However, the literature has proposed multiple interpretations of what precisely constitutes MAD. In one of the first seminal papers on the topic, Du et al.~\cite{18} characterized MAD as an approach where ``\emph{multiple language model instances propose and debate their individual responses and reasoning processes over multiple rounds to arrive at a common final answer}''. Chan et al.~\cite{12} described MAD as a framework where ``\emph{different LLMs can engage in proposing and deliberating unique responses and thought processes across several rounds}'' but do not require that debater agents ``\emph{reach a consensus at the end of the debate}''. Liang et al.~\cite{3} conceptualized MAD as an approach where ``\emph{multiple agents express their arguments in the state of tit for tat and a judge manages the debate process to obtain a final solution}''. 

These definitions illustrate the diversity of perspectives adopted in the literature when conceptualizing MAD. While they share the common idea of multiple AI agents engaging in a deliberative process, they differ with respect to key aspects such as the role of consensus, the structure of agent interactions, and the presence of dedicated coordination or judging mechanisms. As a result, there is currently no widely accepted definition of MAD, nor a clear consensus regarding its essential characteristics, boundaries, and distinguishing features.
Therefore, it is necessary to precisely define what qualifies (and what does not qualify) as MAD within the scope of this systematic study. 

Based on the aforementioned definitions and studies, Table~\ref{tab:mad_criteria} presents the criteria we used to operationalize the concept and establish the boundaries of what constitutes MAD within the scope of this study.

\begin{table}[ht]
\centering
\caption{Definition criteria for Multi-Agent Debate (MAD)}
\label{tab:mad_criteria}
\begin{tabular}{p{0.47\linewidth} p{0.47\linewidth}}
\hline
\textbf{Qualifies as MAD} & \textbf{Does NOT Qualify as MAD} \\
\hline
A discussion between two or more AI agents. 
& Self-reflection by a single agent. \\
\addlinespace
Agents can put forward opposing arguments during the discussion in multiple rounds. 
& One-directional feedback between agents without actual discussion. \\
\addlinespace
AI agents collaboratively discuss to solve the same task. 
& AI agents performing distinct subtasks of the same goal through simple information passing, without active discussions. \\ 
\hline
\end{tabular}
\end{table}

Beyond establishing a clear definition of MAD, we further delimited the scope of this study to approaches applied to natural language tasks. We additionally required debates to occur exclusively among AI agents, without humans acting as debate participants. This restriction follows from the intended scope of this study: MAD is treated as a mechanism for autonomous multi-agent reasoning, whose behaviour and evaluation criteria are qualitatively different from those of human-AI interaction.

\subsection{Search Strategy}
\label{sec:search-strategy}

Based on Kitchenham's PICO strategy to define search strings~\cite{Kitchenham2007}, we designed a search string that operationalizes two key dimensions: the Population (LLMs or AI agents) and the Intervention (Multi-Agent Debate). For each  concept, we included both singular and plural forms and their standard acronyms. The final search string was written as: 

\begin{framed}
\noindent (LLM OR ``Large Language Model'' OR LLMS OR ``Large Language Models'' OR ``AI Agent'' OR ``AI Agents'') AND (MAD OR ``Multi-Agent Debate'')
\end{framed}

The search was conducted using the Scopus database, over the title, abstract, and keywords of papers; no restriction was set in terms of domain or date of publication. 

It is important to clarify that the goal of this initial search was not to retrieve all existing publications on MAD but rather to identify a relevant set of high-quality papers to be used as the seed for the subsequent snowballing process. Accordingly, we prioritized precision over recall. For this reason, we deliberately avoided incorporating additional alternative terms for ``MAD'' or ``Multi-Agent Debate'', since broader terms such as ``debate'' or ``discussion'' introduced substantial noise with very limited relevant results. Finally, we selected Scopus for its broad coverage of high-quality, peer-reviewed research across disciplines~\cite{Martín-Martín2021}. 

\subsection{Study Selection Criteria and Selection Process}
\label{sec:selection-criteria}

Based on the MAD definition and scope described in Section~\ref{sec:mad-definition}, we established the inclusion and exclusion criteria as presented in Table~\ref{tab:criteria}. Notably, we deliberately included MAD strategies from multiple domains rather than restricting our scope to Software Engineering.

The selection process was conducted in three stages: screening by title, followed by abstract review, and detailed full-text assessment. All candidate papers were independently evaluated by two distinct authors, and disagreements were resolved through discussion and consensus.

When multiple versions of the same paper were available on arXiv, the most recent version was selected. In cases where a paper was available both on arXiv and as a conference or journal publication, the latter was prioritized for analysis.

\begin{table}[ht] \centering \caption{Inclusion and Exclusion Criteria} \label{tab:criteria} \renewcommand{\arraystretch}{1.2} \begin{tabular}{p{0.12\linewidth} p{0.83\linewidth}} \toprule \textbf{ID} & \textbf{Criterion} \\ \midrule IC1 & Studies that propose or use one or more debate strategies that satisfy our operationalized definition of MAD. \\ \midrule EC1 & The paper does not provide sufficient details about the debate strategy (e.g., the debate is not the focus of the paper). \\ EC2 & Debates are not exclusively among AI agents. \\ EC3 & Debates are not applied to natural language tasks (e.g., image recognition tasks). \\ EC4 & The paper is not written in English. \\ EC5 & The full text of the paper is not accessible. \\ EC6 & Duplicate or superseded version of the paper.
\\ \bottomrule \end{tabular} \end{table}

All retrieved papers were randomly assigned to two authors of the paper acting as reviewers. The entire selection process was conducted independently, with each reviewer assessing their assigned studies against the predefined inclusion and exclusion criteria. Any disagreements were discussed and resolved through consensus.

\subsection{Snowballing Procedure}
\label{sec:snowballing}

The resulting list of papers served as the seed set for the snowballing process. We conducted a single iteration for both forward and backward snowballing, which helped us to achieve two goals: (1) capturing not only the papers satisfying our query, but also broader relevant works that meet our inclusion/exclusion criteria; and (2) capturing relevant papers available only on arXiv, which today is a fundamental source of timely research in the AI domain that cannot be disregarded in a literature study.

\subsection{Search and Selection Results}
\label{sec:search-results}

\begin{figure}[t]
    \centering
    \includegraphics[width=\linewidth]{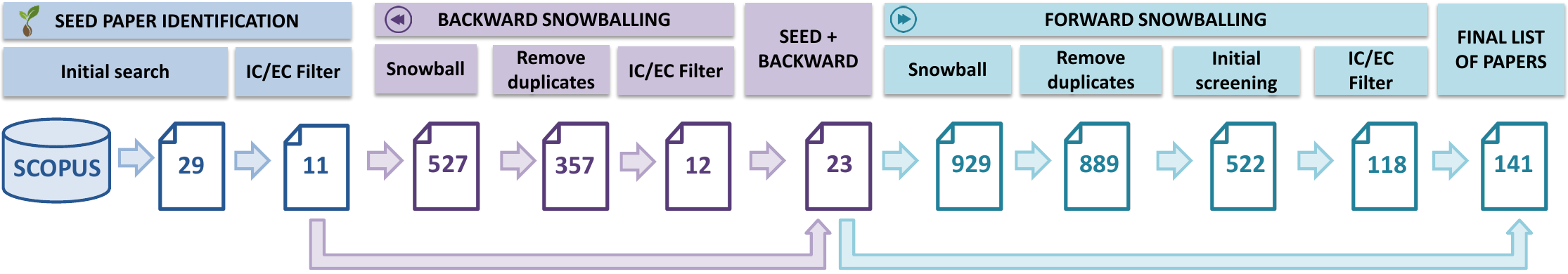}
    \caption{Overview of the search and snowballing process}
    \label{fig:selection-process}
    \Description{Overview of the search and snowballing process, structured into three main stages: (1) seed paper identification; (2) backward snowballing; and (3) forward snowballing. Stage~1 led to 11 studies after inclusion and exclusion criteria analysis. Stage~2 led to 23 studies. Finally, Stage~3 led to 141 studies.}
\end{figure}

Figure~\ref{fig:selection-process} summarizes the complete search and selection process. The search string executed in Scopus yielded 29 results. These were independently reviewed by two authors. Of these, 18 papers either did not meet the inclusion criteria or met one or more exclusion criteria, resulting in 11 seed papers. %No disagreements were observed between the two reviewers at this stage. 

Subsequently, we applied backward snowballing, identifying 527 additional papers. After removing 170 duplicates, 357 papers remained for screening. This set was again independently reviewed by two authors. Any disagreements were resolved through discussion until consensus was reached. Following this process, 12 additional papers satisfied the inclusion/exclusion criteria, bringing the total to 23 papers.

Forward snowballing was subsequently conducted, yielding 929 additional papers. After duplicate removal, 889 papers remained. Given the high number of irrelevant results, an initial filtering step was performed by one author to exclude clearly out-of-scope papers, removing 367 papers and leaving 522 for further review. These papers were then distributed among three of the authors in pairs, ensuring that each paper was reviewed by two authors. Disagreements were discussed and resolved collaboratively. This process resulted in the inclusion of 118 additional papers, leading to a final set of 141 papers. Inter-rater agreement across all screening rounds (initial search, backward and forward snowballing) yielded an average Cohen's kappa of 0.70.

The initial search and backward snowballing were executed in March 2025 and the forward snowballing in July 2025. Among the 141 selected publications, the earliest papers appeared in 2023 (14 papers), followed by 57 publications in 2024 and 70 in 2025, indicating a clear and increasing trend in research activity on this topic.

Among the selected studies, 85 were published in conferences, 10 in journals, 1 as a book chapter, and 45 were available as arXiv preprints.

Applying feature extraction to the selected papers, we identified a number of codes that characterize the different methodologies used in MAD. These codes fall into four primary categories: the \textit{research landscape} with the application domains and tasks where MAD was used, the \textit{participants} involved, the \textit{interactions} held during the debate, and the strategies for reaching \textit{agreement}. Below, we describe how the papers in our review implement MAD in accordance with the defined taxonomy. It is important to note that certain primary studies proposed or evaluated multiple distinct architectural configurations; consequently, our quantitative analysis evaluates a total of 151 distinct MAD approaches derived from the 141 selected papers. The specific codes within these categories are presented in detail throughout the subsequent sections, while the complete data extraction dataset is provided in our replication package.

%\section{Analysis}
%\label{sec:search-analysis}

\section{Research Landscape}
\label{sec:landscape}
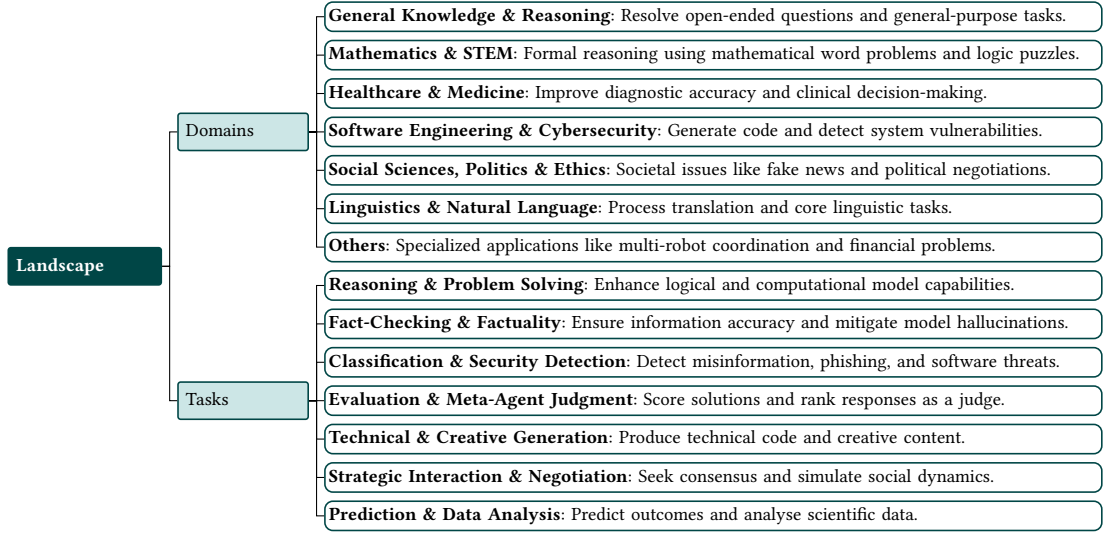
\begin{figure*}[h]
    \centering
    \begin{forest}
    taxonomy tree
    [{\textbf{Landscape}}, rootbox
      [Domains, levelonebox
        [{\textbf{General Knowledge \& Reasoning}: Resolve open-ended questions and general-purpose tasks.}, refswide],
        [{\textbf{Mathematics \& STEM}: Formal reasoning using mathematical word problems and logic puzzles.}, refswide],
        [{\textbf{Healthcare \& Medicine}: Improve diagnostic accuracy and clinical decision-making.}, refswide],
        [{\textbf{Software Engineering \& Cybersecurity}: Generate code and detect system vulnerabilities.}, refswide],
        [{\textbf{Social Sciences, Politics \& Ethics}: Societal issues like fake news and political negotiations.}, refswide],
        [{\textbf{Linguistics \& Natural Language}: Process translation and core linguistic tasks.}, refswide],
        [{\textbf{Others}: Specialized applications like multi-robot coordination and financial problems.}, refswide]
      ]
      [Tasks, levelonebox
        [{\textbf{Reasoning \& Problem Solving}: Enhance logical and computational model capabilities.}, refswide],
        [{\textbf{Fact-Checking \& Factuality}: Ensure information accuracy and mitigate model hallucinations.}, refswide],
        [{\textbf{Classification \& Security Detection}: Detect misinformation, phishing, and software threats.}, refswide],
        [{\textbf{Evaluation \& Meta-Agent Judgment}: Score solutions and rank responses as a judge.}, refswide],
        [{\textbf{Technical \& Creative Generation}: Produce technical code and creative content.}, refswide],
        [{\textbf{Strategic Interaction \& Negotiation}: Seek consensus and simulate social dynamics.}, refswide],
        [{\textbf{Prediction \& Data Analysis}: Predict outcomes and analyse scientific data.}, refswide]
      ]
    ]
    \end{forest}
    \caption{Taxonomy of the multi-agent debate landscape: application domains and addressed tasks.}
    \label{fig:mad-landscape-tree}
    \Description{Taxonomy of the multi-agent debate landscape: application domains and addressed tasks.}
\end{figure*}

The following section analyses the selected literature according to application domains --- the fields of knowledge where MAD is implemented --- and the addressed tasks, which define the specific goals the agents are intended to achieve. These domains and tasks have been derived through an inductive coding process, emerging from a systematic grouping of the analysed corpus based on shared functional and thematic characteristics. The resulting categories, accompanied by concise descriptions\footnote{For each taxonomy value, the text cites representative studies rather than exhaustive lists, since the large number of primary studies --- each potentially mapping to multiple values per dimension --- makes full in-text tracing impractical. Full study-to-value traceability is provided in our replication package.}, are illustrated in Figure~\ref{fig:mad-landscape-tree}. 
This overview illustrates the versatility of MAD across various disciplines and helps categorize the diverse problem sets tackled in recent research.

\subsection{Application Domains}
Application domains refer to the specific fields or subject areas where MAD frameworks are deployed to evaluate or enhance the performance of LLM-based multi-agent systems. % The distribution of the literature analysed across these domains is summarized in Table~\ref{tab:classification_domains}. 
The research landscape spans from general-purpose reasoning to highly specialized professional sectors.

\begin{itemize}
    \item \textbf{General Knowledge \& Reasoning:} This remains the dominant domain, where MAD is used to benchmark the ability of agents to resolve open-ended questions or general-purpose tasks~\cite{12,35,101,124,128,16}. For instance, studies such as Fang et al.~\cite{1} and Wang et al.~\cite{8} use general knowledge Q\&A to measure the reduction of hallucinations and the quality of meta-judgment. This domain serves as the primary ``testing ground'' for new interaction protocols due to the availability of massive, diverse datasets. Commonly utilized benchmarks in this domain include MMLU (Massive Multitask Language Understanding)~\cite{hendrycks2021-MMLU}, BoolQ~\cite{clark-etal-2019-boolq}, BBH (Big-Bench Hard)~\cite{suzgun2022-BBH}, and HotpotQA~\cite{yang-etal-2018-hotpotqa}.

    \item \textbf{Mathematics \& STEM:} A significant portion of research focuses on formal reasoning. Frameworks utilize mathematical benchmarks and logic puzzles to test if collective deliberation can overcome individual calculation errors~\cite{2,3,17,45,23,24}. This domain also includes broader scientific and multidisciplinary reasoning~\cite{18,28,17,164,121,35,40,11}. Standard benchmarks and puzzles used to evaluate these capabilities include GSM8K~\cite{cobbe2021trainingverifierssolvemath} and MATH~\cite{hendrycks2021measuringmathematicalproblemsolving} for arithmetic, CLUTRR~\cite{sinha-etal-2019-clutrr} for logical puzzles, and SCIQ~\cite{welbl2017crowdsourcingmultiplechoicescience} or ARC~\cite{clark2018thinksolvedquestionanswering} for scientific reasoning.

    \item \textbf{Healthcare \& Medicine:} MAD is increasingly applied to clinical settings. Research explores how multi-agent conversations can mitigate cognitive biases in diagnostic reasoning~\cite{15,64,70}, improve medical Q\&A performance~\cite{32,8,37,15,64,70}, and model electronic health records (EHR) for mortality prediction~\cite{44}.

    \item \textbf{Software Engineering \& Cybersecurity:} Applications in this domain focus on the technical reliability of software artifacts. This includes code generation~\cite{11, 26, 29, 30, 35, 52, 127}, code summarization~\cite{12, 88}, and code translation~\cite{88, 107}. In the security sector, frameworks are applied to detect phishing attempts in emails or websites~\cite{57,122} and perform software vulnerability assessments~\cite{67,95}.

    \item \textbf{Social Sciences, Politics \& Ethics:} This domain focuses on societal issues and value alignment. Key research areas include fake news and rumor detection~\cite{9,39,50,58}, political coalition negotiations~\cite{109}, and the mitigation of bias or harmful content in model outputs~\cite{51,131,70,133,61,126}.

    \item \textbf{Linguistics \& Natural Language:} Beyond general reasoning, specific linguistic tasks are addressed through MAD. These include event extraction~\cite{10,108}, machine translation evaluation~\cite{107}, summary source alignment~\cite{76}, and specialized generation tasks like Chinese couplet generation and evaluation~\cite{120}.
    
    \item \textbf{Others:} This category encompasses specialized applications in niche or emerging fields. It includes robotics and navigation tasks such as the coordination of multi-robot aggregation~\cite{19}. The domain also covers consumer and legal services, including decision-making support for products~\cite{43}, customer service speech improvement~\cite{33}, and legal argument generation~\cite{106}. Finally, it incorporates experimental and financial reasoning, such as solving financial problems~\cite{48} and determining correct answers in difficult long-context scenarios or zero-sum games~\cite{124}.
\end{itemize}

From a design perspective, the choice of domain necessitates tailored agent profiles; specialized fields like medicine or law often require agents with specific personas or access to domain-specific knowledge.
From an evaluation standpoint, these domains offer varying degrees of ground-truth clarity: while mathematics provides binary correctness, social and linguistic domains rely on subjective, multi-dimensional metrics.

\subsection{Tasks}
Addressed tasks define the specific functional goals that agents are intended to achieve through debate or collaboration. 
The analysed corpus reveals a diverse range of objectives, which have been classified and grouped into the following categories:
%The analysed corpus reveals a diverse range of objectives, which are summarized and classified in Table~\ref{tab:classification_tasks}. These goals are grouped into the following categories:

\begin{itemize} 
    \item \textbf{Reasoning \& Problem Solving:} This is the most frequent category, focusing on enhancing the logical and computational capabilities of models. It includes arithmetic and mathematical reasoning~\cite{2, 3, 7, 11, 14, 17, 18, 23, 24, 26, 29, 30, 38, 40, 45, 48, 49, 52, 54, 55, 60, 62, 63, 66, 72, 74, 99}, symbolic and common-sense logic~\cite{24, 72}, and deductive reasoning~\cite{118}. In specialized fields, it encompasses medical diagnostic reasoning~\cite{15, 64, 70}.
    
    \item \textbf{Fact-Checking \& Factuality:} These tasks aim to ensure information accuracy and mitigate hallucinations. They include fact verification and retrieval-augmented reasoning~\cite{8, 37, 62}, knowledge graph triple verification~\cite{82}, and the systematic elimination of hallucinations in large language models~\cite{1, 105, 136}.
    
    \item \textbf{Classification \& Security Detection:} This area focuses on content integrity and cybersecurity. Key tasks include the detection of fake news and rumors~\cite{9, 39, 50, 58}, the identification of spam and phishing attempts~\cite{5, 57, 122}, and the assessment of software or smart contract vulnerabilities~\cite{67, 95}. It also covers hate speech detection~\cite{126} and safety classification~\cite{51, 131}.
    
    \item \textbf{Evaluation \& Meta-Agent Judgment:} These frameworks utilize debate to position agents as evaluators (\emph{LLM-as-a-judge}). Tasks include scoring summaries~\cite{56, 65, 76, 101}, evaluating machine translation quality~\cite{107}, and establishing rankings among multiple model responses~\cite{20, 35, 91}.
    
    \item \textbf{Technical \& Creative Generation:} Agents collaborate to produce or refine complex content. This includes code generation~\cite{11, 26, 29, 30, 35, 52, 127} and summarization~\cite{88}, creative tasks like personalized Chinese couplet generation~\cite{120}, and the production of legal arguments~\cite{106}.
    
    \item \textbf{Strategic Interaction \& Negotiation:} This category explores how agents reach agreements or simulate social dynamics to achieve a shared goal. Tasks include consensus seeking~\cite{19, 21, 66}, political coalition negotiations~\cite{109}, cross-cultural negotiation~\cite{27}, and the study of group conformity~\cite{31, 125}.
    
    \item \textbf{Prediction \& Data Analysis:} Advanced analytical applications include modeling electronic health records (EHR) to predict mortality or readmission~\cite{44}, project duplication detection~\cite{121}, and comparative analysis of scientific papers to determine novelty~\cite{164}.
\end{itemize}

From a design perspective, the diversity of addressed tasks necessitates a strategic mapping between task requirements and multi-agent architectures. Objective-driven tasks, such as mathematical problem solving~\cite{2, 3, 17, 23, 48, 49} or code generation~\cite{26, 29, 30, 127}, often rely on iterative refinement through homogeneous agents to converge on a single correct answer. In contrast, knowledge-intensive and subjective tasks benefit from the design of heterogeneous personas~\cite{25, 59, 164} that ensure a broad coverage of perspectives and help mitigate individual cognitive biases, particularly in medical contexts~\cite{15, 64, 70}. Furthermore, security-critical tasks like vulnerability assessment~\cite{67, 95} or phishing detection~\cite{57, 122} require the implementation of adversarial or ``red-teaming'' protocols~\cite{131} to effectively probe for system weaknesses.

From an evaluation standpoint, the variety of functional goals implies that no single metric can capture performance across all domains. While arithmetic and logical tasks allow for objective measurement through accuracy scores on established benchmarks like GSM8K~\cite{45}, open-ended and creative tasks increasingly rely on meta-evaluation frameworks where agents act as judges to score coherence, faithfulness, and alignment~\cite{12, 35, 56, 65, 76, 91, 101}. This shift towards multi-agent evaluation introduces new challenges, such as the risk of inter-agent sycophancy~\cite{45} or the reinforcement of existing biases, which must be carefully monitored through diverse reasoning datasets and multi-dimensional scoring systems.

\begin{rqfindings}{1}
\begin{itemize}
    \item[\faHandPointRight] \textbf{MAD is primarily a general-purpose reasoning paradigm, not yet a domain-specific methodology}. Most studies use MAD in general knowledge, mathematical, or formal reasoning tasks, in which correctness is easier to benchmark and debate gains are easier to quantify~\cite{2,3,12,17,18,24,35,45,49,124,128}.

    \item[\faHandPointRight] \textbf{Applied domains use MAD as a reliability layer}. In medicine, software engineering, cybersecurity, fake news detection, and safety-related tasks, debate is mainly introduced to reduce errors, expose weaknesses, or improve trustworthiness rather than to model deliberation for its own sake~\cite{15,50,57,64,67,70,95,122,131}.

    \item[\faHandPointRight] \textbf{Evaluation maturity varies strongly by task type}. Objective tasks such as mathematics, logic, and code generation support direct benchmark-oriented evaluation, while open-ended tasks such as text evaluation, legal reasoning, negotiation, and creative generation rely on weaker proxies, including \emph{LLM-as-a-judge} and qualitative synthesis~\cite{12,35,56,76,91,101,106,107,109,120}.
\end{itemize}
\end{rqfindings}

\section{Participants}
\label{sec:participants}

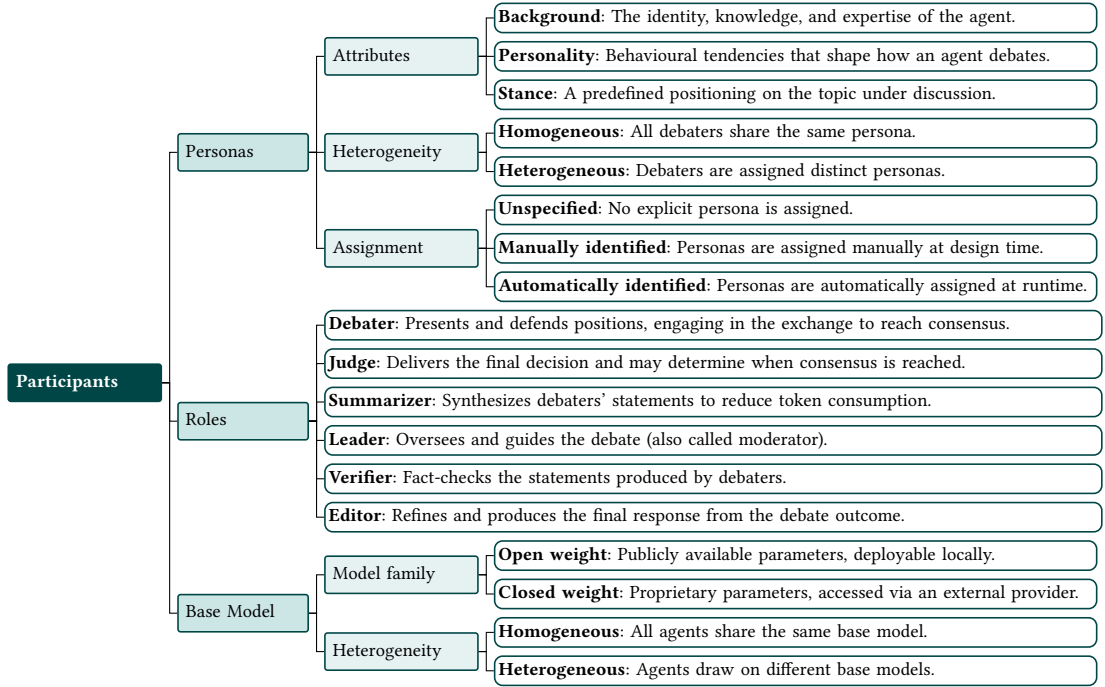
\begin{figure*}[ht]
    \centering
    \begin{forest}
    taxonomy tree
    [{\textbf{Participants}}, rootbox
      [Personas, levelonebox
        [Attributes, leveltwobox
          [{\textbf{Background}: The identity, knowledge, and expertise of the agent.}, refs]
          [{\textbf{Personality}: Behavioural tendencies that shape how an agent debates.}, refs]
          [{\textbf{Stance}: A predefined positioning on the topic under discussion.}, refs]
        ]
        [Heterogeneity, leveltwobox
          [{\textbf{Homogeneous}: All debaters share the same persona.}, refs]
          [{\textbf{Heterogeneous}: Debaters are assigned distinct personas.}, refs]
        ]
        [Assignment, leveltwobox
          [{\textbf{Unspecified}: No explicit persona is assigned.}, refs]
          [{\textbf{Manually identified}: Personas are assigned manually at design time.}, refs]
          [{\textbf{Automatically identified}: Personas are automatically assigned at runtime.}, refs]
        ]
      ]
      [Roles, levelonebox
        [{\textbf{Debater}: Presents and defends positions, engaging in the exchange to reach consensus.}, refswide]
        [{\textbf{Judge}: Delivers the final decision and may determine when consensus is reached.}, refswide]
        [{\textbf{Summarizer}: Synthesizes debaters' statements to reduce token consumption.}, refswide]
        [{\textbf{Leader}: Oversees and guides the debate (also called moderator).}, refswide]
        [{\textbf{Verifier}: Fact-checks the statements produced by debaters.}, refswide]
        [{\textbf{Editor}: Refines and produces the final response from the debate outcome.}, refswide]
        %[{\textbf{HITL}: A human participant is involved in the debate.}, refswide]
      ]
      [Base Model, levelonebox
        [Model family, leveltwobox
          [{\textbf{Open weight}: Publicly available parameters, deployable locally.}, refs]
          [{\textbf{Closed weight}: Proprietary parameters, accessed via an external provider.}, refs]
        ]
        [Heterogeneity, leveltwobox
          [{\textbf{Homogeneous}: All agents share the same base model.}, refs]
          [{\textbf{Heterogeneous}: Agents draw on different base models.}, refs]
        ]
      ]      
    ]
    \end{forest}
    \caption{Taxonomy of participants in multi-agent debate systems.}
    \label{fig:participants-taxonomy}
    \Description{Taxonomy of participants in multi-agent debate systems.}
\end{figure*}

The following section examines how participants are designed in MAD frameworks, categorized into three main dimensions summarized in Figure~\ref{fig:participants-taxonomy}: the \textit{Personas} assigned to participants, the functional \textit{Roles} they occupy, and the \textit{Base Model} underlying their reasoning capabilities.

\subsection{Personas}

\begin{definition}{Personas}{def:personas}
The set of individual characteristics, independent of a participant's functional role, that shape how the agent reasons and argues during the debate.
\end{definition}

Formally, a persona may be assigned to each participant $v \in V$, where $V$ denotes the set of participants in the debate. 
Beyond what its functional role determines, the persona conditions the viewpoint an agent brings to the exchange, and thereby shapes the diversity of the deliberation. 
In practice, personas are instantiated through the agent's system prompt, which may specify different kinds of persona-related information, such as the agent's background, personality traits, and stance towards the debate topic. 
We characterize personas along three descriptors: their \emph{heterogeneity} across participants, how they are \emph{assigned}, and the \emph{attributes} through which they are expressed, which we categorize into three families: background, personality, and stance.
Any of these information families may be specified explicitly or left unspecified. 

\subsubsection{Attributes}

Personas are specified through a set of attributes that shape how an agent behaves or interacts with other agents: a \emph{background} that situates what the agent knows, a \emph{personality} of behavioural traits that shape how it argues, and a \emph{stance} that fixes the position it must defend.

A \emph{background} attribute models who an agent is: its identity, domain knowledge, expertise, skills, role, or demographic context. It grounds what the agent is presumed to know and the point of view it argues from, without committing it to a position on the topic of the debate. 
Backgrounds range from specialized functional roles to richer social or cultural identities. PhishDebate~\cite{122} instantiates dedicated inspectors (HTML, URL, content, and brand agents).  
CodeGen~\cite{127} defines distinct coding roles (e.g., \texttt{PythonAssistant}, \texttt{AlgorithmDeveloper}).
Chen et al.~\cite{124} pair expert debaters with a non-expert judge to vary expertise. 
Other works ground the background in cultural or demographic context, as in the culturally differentiated agents of Zhang et al.~\cite{27} or the political profiles of Moghimifar et al.~\cite{109}.

A \emph{personality} or behavioural attribute modulates how an agent argues, rather than the position it defends. 
Chen et al.~\cite{19} show that stubborn and suggestible traits affect consensus dynamics, with stubborn agents exerting stronger influence on the final value; related work explores easy-going and overconfident profiles~\cite{76}, and DEBATE varies a critic's degree of criticality from weakly to strictly negative~\cite{56}.
A neighbouring pattern is role-play, assigning agents distinct discussion personality traits through structured models such as the Six Thinking Hats in LLM discussion~\cite{92}. 
These attributes offer a finer-grained lever over debate dynamics, diversity, and convergence, but their effects are harder to isolate than stance.

A \emph{stance} assigns an agent a predefined position. 
In two-sided settings, it is often implemented by assigning opposing positions: devil's-advocate or critic roles that challenge another agent's output~\cite{56}, prosecution and defence arguing whether a news item is fake or genuine~\cite{50}, faithful and unfaithful evaluators in summary-faithfulness assessment~\cite{76}, proponent and challenger in retrieval debate~\cite{136}, or supporting and opposing agents in stance detection~\cite{75}.
Other formulations tie the stance to the answer space, as in the gold-versus-distractor setup of Chen et al.~\cite{124}, where one debater argues for the correct answer and the other for the strongest incorrect option. 
Stance assignment may itself be manipulated: MADISSE~\cite{76} initializes uniformly distributed faithful/unfaithful stances, while Courtroom-FND~\cite{50} switches prosecution and defence across rounds. 
Predefined stances make disagreement explicit and provide a controllable adversarial structure, though they may impose conflict even when the task would not naturally require it.

\subsubsection{Heterogeneity}

Heterogeneity refers to whether participants operate under a single shared persona or several distinct ones.
A \emph{homogeneous} configuration (employed by 44.4\% of the surveyed approaches\footnote{All percentages are computed over the set of distinct MAD approaches ($n=151$). For a given property, totals may be below 100\% when some approaches leave that property unspecified and it cannot be inferred from the primary study without assumptions. Conversely, totals may exceed 100\% when a single approach combines or compares several strategies and is therefore counted under more than one value.}) assigns one persona to all participants $v \in V$, so they argue from a common perspective and differ only in the arguments they generate~\cite{1,2,18,52,55,60,74,78,99,105,128}. 
This setting covers both the canonical multi-debater setup --- several instances of the same agent revising answers under an identical prompt --- and designs that replicate a single specialized persona across participants. 
Holding the persona fixed deliberately isolates the effect of other design choices.
As an example, Wang et al.~\cite{37} vary only the collaboration mechanism (governance, participation, interaction pattern, and context management). 
Similarly, GroupDebate~\cite{81} restructures the communication topology, both under a shared persona. 
When the contrast is made explicit, the homogeneous case typically serves as a baseline against persona diversity, as in ChatEval~\cite{12}. 
A shared persona simplifies implementation and isolates the interaction mechanism, but constrains reasoning diversity, leaving agents prone to premature convergence and shared errors.

A \emph{heterogeneous} configuration (53.6\%) assigns distinct personas so that at least two participants reason from different perspectives~\cite{3,11,25,40,43,62,95,122,163}. This is realized through any of the persona attributes discussed above: opposing stances such as the ``angel'' and ``devil'' debaters~\cite{3,11}, differentiated backgrounds as in PhishDebate~\cite{122}, panels of domain experts~\cite{62,95}, or automatically generated agent sets~\cite{25,43,163}. 
Notably, Liu et al.~\cite{40} find that distinct personas alone may not elicit genuinely divergent reasoning, and pair them with distinct reasoning strategies to break the agents' shared ``mental set''. 
Heterogeneous personas are the primary mechanism through which MAD operationalizes perspective diversity and mitigates conformity, at the cost of additional variability, since outcomes become sensitive to which personas are assigned and how they are specified.

\subsubsection{Assignment}

Assignment captures whether and how explicit personas are established. 
The relevant characteristics --- expertise, domain knowledge, stances, or behavioural traits --- may be left unspecified, defined manually at design time, or generated automatically at runtime.

An \emph{unspecified} assignment (53.6\%) provides no persona beyond the generic task prompt~\cite{2,4,7,16,18,20,21,23,24}. 
Participants may still act as debaters, judges, or other functional roles (see Section~\ref{sec:roles}), but carry no distinct expertise, viewpoint, or behavioural identity. 
This differs from a homogeneous configuration, which explicitly assigns the \emph{same} persona to all participants rather than leaving it undefined.

A \emph{manually identified} assignment (28.5\%) fixes a predetermined panel before the debate~\cite{27,30,31,48,52,54,62,64,106,117,124,127}. 
For instance, Ma~\cite{62} casts debaters as historical philosophers in a structured ``symposium''. 
Barbi et al.~\cite{127} assign four complementary, heterogeneous coding personas.
Zhang et al. instantiate adversarial roles such as plaintiff and defendant in legal-argumentation settings~\cite{106}. 
Manual assignment gives precise control over panel composition but requires prior assumptions about which perspectives are relevant to the debate.

An \emph{automatically identified} assignment (17.9\%) generates personas at runtime as a function of the task~\cite{25,26,29,43,87,92,163,164}. 
Wang et al.~\cite{25} prompt a single model to instantiate a task-dependent set of personas. 
Song et al.~\cite{29} identify the roles required per subtask and assemble a matching team.
Similarly, ChoiceMates~\cite{43} spawns three to six agents per query, while town-hall-style prompting generates a persona list adapted to the task~\cite{163}.
Automatic assignment improves adaptability and reduces manual effort, but makes debate quality dependent on the persona-generation step.

\subsection{Roles}
\label{sec:roles}

\begin{definition}{Roles}{def:roles}
Roles refer to the specific functions, responsibilities, or behavioural profiles assigned to participants in order to structure the interaction and guide the debate process.
\end{definition}

The primary participant is the \textit{Debater}: an agent that presents arguments, reacts to others, and contributes according to its prompt, stance, persona, or task-specific expertise. 
Debaters may be generic agents exchanging candidate answers or rationales~\cite{2,5,7,12,17,18,19,20,21,24}, or specialized participants, such as proponent/opponent agents in jailbreak detection~\cite{151}, technical expert agents in phishing detection~\cite{122}, or advocate/skeptic agents in annotation-oriented summarization~\cite{161}. 
Beyond debaters, we identify the following recurring roles:

\begin{itemize}
    \item \textbf{Judge} (\emph{38.4\%}): A single agent produces the final decision --- selecting a winning answer, aggregating arguments, or determining whether consensus is reached~\cite{1,3,4,9,10,11,15,16,22,35,37,56,121,124,131,151,161}. This role operationalizes the \emph{LLM-as-a-judge} paradigm~\cite{107,154}, in which an LLM is prompted to deliver human-like assessments of candidate outputs. 
    CFMAD~\cite{1} uses a judge to evaluate the debate after agents argue from counterfactual stances.
    Auto-Arena~\cite{35} relies on a committee of judges deciding by majority vote.

    \item \textbf{Summarizer} (\emph{21.9\%}): Synthesizes debate statements to reduce context length or provide a compact record for decision-making~\cite{4,8,9,10,15,16,29,31,36,37,44,46,120,131,146}. In ERD~\cite{4} and FORD~\cite{16} the judge also summarizes, while in Song et al.~\cite{29} a reflector LLM summarizes and analyses the dialogue before feeding it back to the Captain Agent.

    \item \textbf{Leader} (\emph{17.2\%}): Also known as moderator, the leader coordinates the debate by managing participation, turns, speakers, or discussion structure~\cite{5,8,15,25,29,37,44,46,64,72,87,91,93,108,118,122,126,145,146,160,164}. Beyond turn-taking, some leaders actively shape the discussion: the moderator in Co-STORM~\cite{87} steers the discourse and updates the participant list as the debate evolves, while in SMoA~\cite{145} it also filters which contributions advance. 

    \item \textbf{Verifier} (\emph{6.0\%}): Also referred to as factuality checker, tester, or auditor~\cite{14,23,93,95,99,111,120,125,127}. It checks the correctness, factuality, or validity of statements and intermediate outputs. For example, LLM-SmartAudit~\cite{95} audits smart-contract vulnerabilities, while CodeGen~\cite{127} instantiates verification as testing generated code.

    \item \textbf{Editor} (\emph{3.3\%}): Refines, composes, or produces the final response from the debate outcome~\cite{11,59,85,114,125}. As an example, Debate-to-Write~\cite{59} turns the outcome into an argument plan, while PEG~\cite{85} uses a generator to produce the final answer from the debate consensus.

    % \item \textbf{Human-in-the-loop (HITL) participant} (\emph{2.6\%}): Some approaches admit humans as active or optional participants rather than only LLM-based agents~\cite{43,87,125,146}. Co-STORM~\cite{87} lets users intervene with questions or arguments, while Plurals~\cite{125} supports optional human participation in customizable deliberation structures.
\end{itemize}

These roles are not mutually exclusive: a single agent may combine judge and summarizer~\cite{4,16}; leader, summarizer, and judge~\cite{37,44}; or moderator and summarizer~\cite{125,146}. Role assignment is also frequently intertwined with persona design, as several approaches assign backgrounds, stances, or behavioural profiles to debaters to ensure that different perspectives are represented~\cite{5,9,12,14,15,31,39,56,76,92,122,161}.

\subsection{Base Model}

\begin{definition}{Base Model}{def:base-model}
The underlying pre-trained LLM(s) used by the agents participating in the debate.
\end{definition}

Base models can be characterized along several axes --- scale (e.g., small vs.\ large), architecture, or training paradigm, among others. We focus on two descriptors: their \emph{model family} (open- vs.\ closed-weight) and their \emph{heterogeneity} across agents within a debate. We adopt the open/closed distinction because it is objective and stable --- a model's weights are either publicly accessible or not --- whereas alternatives like ``small'' vs.\ ``large'' lack consensus and shift as the field evolves. It is also the most consequential for debate design, governing whether agents run locally or through external providers, with implications for cost, reproducibility, and control.

\subsubsection{Model family}

We distinguish \emph{open-weight} models, whose parameters are publicly accessible and locally deployable, from \emph{closed-weight} models, which are proprietary and accessed through external providers.

Among the surveyed approaches, 66 rely exclusively on closed-weight models (43.7\%), 32 exclusively on open-weight models (21.2\%), and 53 combine both (35.1\%). Closed-weight configurations are typically based on GPT-family models, sometimes combined with Claude, Gemini, Bard, or PaLM~\cite{1,18,23,26,32,56,96,103,164}, whereas open-weight ones rely on Llama, Qwen, Mistral, Gemma, Vicuna, DeepSeek, or other locally deployable models~\cite{7,27,30,31,55,99,127,149}. Mixed configurations are common in studies comparing proprietary and open-weight models under the same debate protocol~\cite{2,3,29,35,37,54,58,60,138}.

By family, GPT dominates (118 approaches, 78.1\%), followed by Llama (61, 40.4\%), Qwen (28, 18.5\%), Claude (22, 14.6\%), Mistral (20, 13.2\%), Gemini (14, 9.3\%), DeepSeek and Gemma (12 each, 7.9\%), and Vicuna (8, 5.3\%); Yi, Baichuan, Phi, PaLM, SeaLLM, EXAONE, and InternLM appear less frequently. MAD research thus remains strongly centred on GPT, although open-weight alternatives are increasingly used.

\subsubsection{Heterogeneity}

A debate instance is \emph{homogeneous} when all agents share the same base model --- even if the paper reports separate experiments with several models --- and \emph{heterogeneous} when agents within the same debate draw on different models.
Homogeneous configurations (81.5\%) predominate over heterogeneous settings (27.8\%). They are typically used to isolate the effect of debate structure, prompting, topology, or role assignment while holding the model fixed~\cite{18,19,21,37,56,127,149}; some evaluate several candidate models but run each debate within a single family, preserving homogeneity~\cite{149}.

Heterogeneous configurations introduce model-level diversity, combining proprietary and open-weight models to test whether differing capabilities or inductive biases improve outcomes~\cite{22,33,35,37,54,58,60,96,103,138}. Heterogeneity is sometimes explicit in the design rationale: Debate and Reflect~\cite{54} assigns stronger models to teacher roles and smaller ones elsewhere, and Debate-to-Detect~\cite{58} reports mixed-model configurations outperforming homogeneous setups in misinformation detection. Such diversity may increase robustness, but it complicates attribution, since gains may stem from model capability rather than the debate protocol alone.

\begin{rqfindings}{2}
\begin{itemize}
    \item[\faHandPointRight] \textbf{Persona diversity is recognized but rarely operationalized systematically}. When diversity is used, it is typically introduced through manually assigned roles, stances, or expertise profiles rather than evaluated as an independent design variable, making it hard to determine which forms of diversity actually improve debate quality~\cite{12,25,37,40,62,95,122,163}.

    \item[\faHandPointRight] \textbf{Stance is a more mature control mechanism than personality}. Stance assignment is a practical way to force disagreement and surface counterarguments, especially through adversarial pairings such as angel/devil, faithful/unfaithful, prosecution/defence, or proponent/challenger~\cite{3,11,50,56,75,76,124,136}. Personality traits are less consistently defined and harder to isolate experimentally, though early evidence suggests that stubbornness, suggestibility, confidence, or criticality can substantially affect convergence~\cite{19,56,76,92}.

    \item[\faHandPointRight] \textbf{Judge and summarizer roles signal a shift from pure debate to meta-coordination}. Many approaches add judges, summarizers, leaders, or hybrid meta-agents to control, compress, and resolve the debate rather than relying only on peer-to-peer deliberation~\cite{1,4,16,29,35,37,44,56,124,146}. MAD performance thus increasingly depends not only on how debaters argue, but on how coordination roles curate information flow and turn debate traces into final outputs.

    \item[\faHandPointRight] \textbf{Model homogeneity remains the default despite the promise of model-level diversity}. Most debates instantiate agents from the same base model, even when studies benchmark several independently. Heterogeneous configurations offer a principled path to greater reasoning diversity, particularly when combining stronger and weaker models or closed- and open-weight families~\cite{22,35,37,54,58,60,96,103,138}; future evaluations should separate the effect of model diversity from that of the debate protocol.

    \item[\faHandPointRight] \textbf{Reproducibility is constrained by closed-weight dependence and underspecified participant design}. MAD research remains strongly centred on GPT-family and other closed-weight models, while participant prompts, personas, role combinations, and model assignments are often described only partially~\cite{1,18,23,26,32,56,96,103,164}. This limits reproducibility and cross-study comparison, making explicit participant specifications essential for future benchmarking.
\end{itemize}
\end{rqfindings}

\section{Interaction}
\label{sec:interaction}

\begin{figure*}[b]
    \centering
    \begin{forest}
    taxonomy tree
    [{\textbf{Interaction}}, rootbox
      [Topology, levelonebox
        [Adaptability, leveltwobox
          [{\textbf{Static}: Communication channels remain unchanged throughout the debate.}, refs],
          [{\textbf{Dynamic}: Communication channels adapt throughout the debate.}, refs]
        ]
        [Structure, leveltwobox
          %[{\textbf{Bilateral}: Two agents exchange arguments directly.}, refs],
          [{\textbf{Fully connected}: All agents interact with each other.}, refs],
          [{\textbf{Divided into groups}: Agents interact within disconnected subgroups.}, refs],
          [{\textbf{Structured networks}: Interaction follows a predefined network topology.}, refs]
        ]
      ]
      [Protocol, levelonebox
        [Mechanism, leveltwobox
          [{\textbf{Sequential}: Agents intervene in ordered turns.}, refs],
          [{\textbf{Simultaneous}: Agents contribute to the debate at the same time.}, refs],
          [{\textbf{Hybrid}: Combines sequential and simultaneous exchanges.}, refs]
        ]
      ]
      [Format, levelonebox
        [Syntax, leveltwobox
          [{\textbf{Natural language}: Arguments are exchanged in human-readable text.}, refs],
          [{\textbf{Embeddings}: Arguments are exchanged as vector embeddings directly.}, refs]
        ]
        [Content, leveltwobox
          [{\textbf{Verbatim}: Full arguments are shared without modification.}, refs],
          [{\textbf{Summarized}: Arguments are compressed into shorter representations.}, refs],
          [{\textbf{Other}: Alternative content transformation strategies are applied.}, refs]
        ]
        [Memory access, leveltwobox
          [{\textbf{Short-Term}: Access limited to current debate context.}, refs],
          [{\textbf{Long-Term}: Access extends to persistent memory across debates.}, refs],
          [{\textbf{Knowledge Retrieval}: Agents query external knowledge sources.}, refs],
          [{\textbf{Parametric}: Debates embedded into model parameters through training.}, refs]
        ]
      ]
    ]
    \end{forest}
    \caption{Taxonomy of interaction in multi-agent debate systems.}
    \Description{Taxonomy tree of the interaction dimension, organized into topology (adaptability, structure), protocol (mechanism), and format (syntax, content, memory access).}
    \label{fig:mad-topology-tree}
\end{figure*}
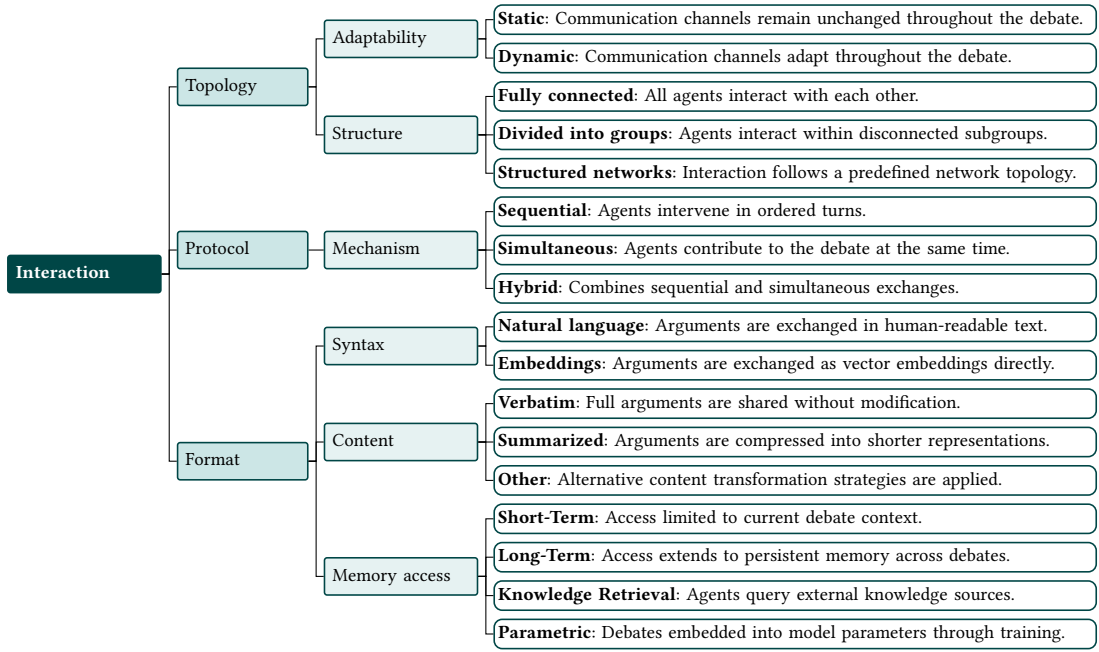

The following section examines how information exchange is structured in MAD frameworks, categorized into three main dimensions summarized in Figure~\ref{fig:mad-topology-tree}: the \textit{Topology} organizing the communication structure, the \textit{Protocol} governing the order and timing of exchanges, and the \textit{Format} defining how arguments are represented and delivered.

\subsection{Topology}

\begin{definition}{Topology}{def:topology}
The structural organization of agents and their communication links in the debate. It determines how agents are connected, grouped, and allowed to exchange information.
\end{definition}

Building on classical graph theory~\cite{diestel2025graph}, we formalize the topology of a MAD system as a directed graph $G = (V, E)$, where $V$ is the set of agents and $E \subseteq V \times V$ the set of communication paths. An edge $(u,v) \in E$ indicates that agent $u$ can transmit information to agent $v$; the path is unidirectional if $(u,v) \in E$ but $(v,u) \notin E$, and bidirectional if both hold. This defines the structural constraints under which agents exchange arguments, characterized by two descriptors: \emph{adaptability} and \emph{structure}.

\subsubsection{Adaptability}

Adaptability refers to whether the communication structure changes during the debate, i.e., whether the edge set $E$ remains fixed or evolves over time.

A \emph{static} interaction (88.7\%) keeps the graph $G = (V, E)$ invariant, reflecting an interaction structure predefined at design time~\cite{3,4,69,70,95,106,116}. The vast majority of MAD approaches are static: all two-agent settings are trivially fixed, and several works deliberately compare alternative fixed strategies.
Chan et al.~\cite{12} define three unchanging communication protocols, and Zhang et al.~\cite{52} model agents as spatial-temporal graphs with connectivity held constant during inference. 
Static schemes simplify implementation, enable controlled experimentation by isolating other variables (prompting, roles, model settings), and yield predictable token consumption that can be estimated in advance.
Their rigidity, however, limits runtime adaptation.

A \emph{dynamic} interaction (9.9\%) lets the edge set evolve, becoming time-dependent $E(t)$, so agents adapt their communication in response to context or intermediate results. Some approaches activate previously disconnected agents: DyLAN~\cite{26} selects agents from a candidate pool and dynamically deactivates them during solving, Song et al.~\cite{29} assemble and revise teams per subtask, and Park et al.~\cite{43} let users invoke or switch among agent subsets. Others keep the agent set fixed but adapt the graph across rounds: Wang et al.~\cite{30} optimize a multi-round graph via \emph{Node Dropout} and \emph{Edge Dropout}, and Sun et al.~\cite{49} update edge weights so agents interact only with beneficial peers. Dynamic interaction improves flexibility and efficiency by focusing communication on the most relevant agents, at the cost of reduced experimental control, since outcomes depend on the update policy for $E(t)$.

\subsubsection{Structure}

Structure characterizes how nodes (agents) are connected and how communication is distributed across $G = (V, E)$.

A \emph{fully connected} structure (72.2\%) links every pair of agents ($\forall u,v \in V,\ (u,v) \in E$), so every message reaches all other agents. The \emph{bilateral} case ($|V| = 2$) is its trivial instance. Most surveyed approaches adopt this structure~\cite{5,17,37,38,58,59,99}.
Combined with a dynamic topology, it can yield reduced fully connected graphs by removing low-performing agents (and their incident edges) while preserving connectivity among the rest~\cite{26}. 
Variability within fully connected designs is otherwise very limited.

A \emph{divided into groups} structure (3.3\%) partitions agents into disjoint subsets $V_1, V_2, \dots, V_k$ with communication restricted mostly within each group, i.e., $(u,v) \in E \Rightarrow \exists i \; (u \in V_i \land v \in V_i)$. Liu et al.~\cite{81} split agents into debate groups, deferring cross-group exchange to stage boundaries via shared summaries.
Wang et al.~\cite{22} let each agent access full answers within its group but only viewpoints from outside it.
Sun et al.~\cite{48} randomly split agents into two teams later adjudicated by a separate judge. 
These partitions balance diversity and coordination: agents deliberate intensively within subsets, while inter-group communication is mediated through summaries, judges, or specialized roles.

\begin{figure}[t]
    \centering
    \includegraphics[width=\linewidth]{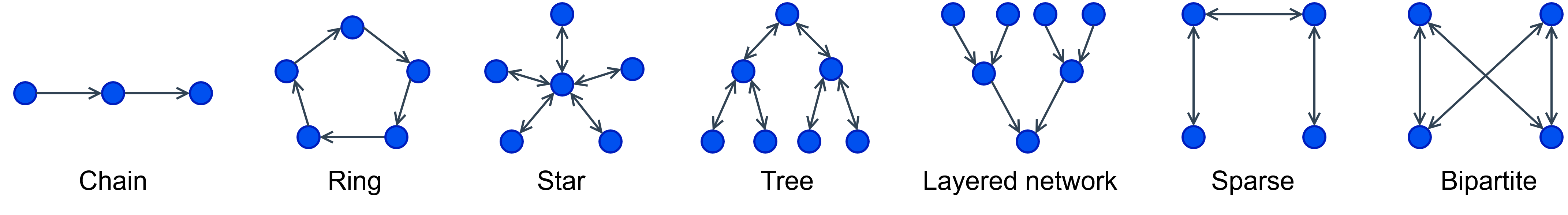}
    \caption{Overview of structured network topologies for interaction in MAD systems.}
    \label{fig:topology-structure}
    \Description{Overview of interaction topology structured networks in MAD systems, identifying: chain, ring, star, tree, layered network, sparse, and bipartite.}
\end{figure}

A \emph{structured network} (23.2\%) constrains $E$ to an explicit graph (e.g., star, chain, tree). As illustrated in Figure~\ref{fig:topology-structure}, and based on the analysed papers, several canonical structures emerge:

\begin{itemize}

\item \textbf{Chain}: a linear topology where $(v_i, v_{i+1}) \in E$, so each agent builds on its predecessor. Huang et al.~\cite{5} introduce sequential modes that accumulate reasoning over time, and others use a chain as a baseline against more advanced networks~\cite{67}. It promotes incremental refinement at low overhead but is prone to error propagation and limits parallel exploration.

\item \textbf{Ring}: a cyclic chain where $(v_i, v_{i+1}) \in E$ and $(v_n, v_1) \in E$, so each agent communicates with a fixed neighbour set. 
Yin et al.'s \emph{Relay} paradigm connects models in a circle, each receiving from its predecessor and passing to its successor~\cite{72}.
Sparse neighbour-connected topologies follow a similar principle~\cite{3}. 
It balances information sharing and efficiency, reducing redundancy relative to full connectivity while retaining iterative refinement.

\item \textbf{Star}: a centralized topology where a single node aggregates from all others, as in M-MAD, where debate groups produce intermediate results synthesized by an agent (e.g., a \emph{judge agent}) into a final decision~\cite{107}. This improves global consistency and interpretability but introduces a central bottleneck and single point of failure.

\item \textbf{Tree}: a hierarchical topology enabling recursive decomposition of reasoning. Tree-of-Debate constructs a dynamic tree whose nodes are subtopics and whose edges refine arguments~\cite{164}, exploring reasoning paths in parallel while preserving structured progression, at the cost of higher coordination complexity.

\item \textbf{Layered network}: a multi-stage topology where agents form sequential layers, each performing a distinct transformation. AgentPrune's pipeline of dimension partition, intra-dimension debate, and final aggregation exemplifies this~\cite{52}, enabling modular reasoning and specialization but limiting cross-layer interaction.

\item \textbf{Sparse}: a low-density topology where each agent interacts with a limited subset of peers ($|E| \ll |V|(|V|-1)$). Li et al.~\cite{2} systematically study sparse communication (e.g., neighbour-connected graphs) and show that reducing connectivity can retain or even improve performance while significantly lowering inference cost, though it requires care to avoid removing critical information flows.

\item \textbf{Bipartite}: agents split into two disjoint sets interacting primarily across the partition, as in M-MAD, where paired agents with opposing stances debate within each dimension~\cite{107}. This enhances contrastive reasoning and role specialization but may limit intra-group sharing.

\end{itemize}

These structures expose a central trade-off: greater connectivity improves information sharing and robustness, while restricting it improves efficiency and scalability. The choice of topology therefore directly shapes the balance among reasoning quality, interpretability, and computational cost.

\subsection{Protocol}

\begin{definition}{Protocol}{def:protocol}
The set of interaction rules that governs how agents participate in the debate. It specifies the order, timing, and coordination of exchanges among agents.
\end{definition}

Building on the graph-based formalization of topology, the protocol is the temporal dimension governing how interactions unfold over $G = (V, E)$. Let $T = \{1, \dots, t_{\max}\}$ denote the discrete debate steps; at each $t \in T$, a communication event transmits a message (a sequence of tokens) from agent $u$ to agent $v$ along an edge $(u,v) \in E$, defining when interactions occur and how information propagates across successive rounds.

\subsubsection{Mechanism}

Mechanism specifies how communication events are scheduled over time, i.e., the subset $E(t) \subseteq E$ of edges instantiated at each step $t$.

A \emph{sequential} mechanism (42.4\%) activates one event per step ($|E(t)| = 1$), enforcing an ordered, linear flow in which each message depends on previously generated content~\cite{116,117,120,121,123,126,127,133,136}. These strategies unfold over a fixed number of rounds in which the agent set is iteratively traversed until consensus is reached or an iteration cap is hit~\cite{31,64,75,124}. Ordering can be instantiated in several ways: randomized turn-taking to mitigate ordering bias and dominance effects~\cite{31,62}, a predefined or dynamically controlled order mediated by a coordinator (e.g., a conversation manager or instructor agent) that selects the next speaker from the evolving context~\cite{29}, or explicit comparison of both strategies~\cite{37}. 
As in human discussion, ordering is non-trivial: early contributions can anchor later reasoning and amplify conformity, making interaction order a key factor in the diversity and robustness of deliberation.

A \emph{simultaneous} mechanism (47.0\%) activates multiple events in parallel ($|E(t)| > 1$), letting agents contribute concurrently under identical context at step $t$~\cite{2,7,9,11,92,103,122,128}.
\emph{Simultaneous talk}, formalized in Wang et al.~\cite{37}, has all agents generate independently within a round and broadcast to peers.
In EvalSVA~\cite{67}, agents perform simultaneous assessment using only shared context from previous rounds. The simplest instance is \emph{bilateral simultaneous debate}~\cite{9,20,38,39,41,50,51}, where two agents argue in parallel and refine across rounds; larger settings extend this to many agents producing responses concurrently before aggregation~\cite{26,40,44,45,65}. Some works directly compare mechanisms: Huang et al.~\cite{5} and Wen et al.~\cite{67} show that simultaneous interaction reduces inter-agent dependency while preserving diversity, and Wang et al.~\cite{37} evaluate its impact on accuracy and token efficiency. By decoupling agent execution within each round, simultaneous mechanisms enable parallel computation and reduce ordering bias, but shift complexity to aggregation, where conflicting or redundant outputs must be reconciled.

A \emph{hybrid} mechanism alternates sequential and simultaneous phases, interleaving independent contributions with interdependent refinement. Though this mechanism is less common (8.0\%), several works adopt such staged pipelines: Wang et al.~\cite{118} explore reasoning branches independently before structured merging, and CollabEval~\cite{65} alternates independent evaluation with collaborative discussion. 
Others operate at finer granularity --- GVIC~\cite{80} interleaves independent reasoning with periodic communication, MAMM-REFINE~\cite{98} combines parallel evaluation with sequential refinement, Ki et al.~\cite{116} switch dynamically between independent and interactive behaviour, and RedDebate~\cite{131} cycles parallel generation and sequential critique --- or extend the idea hierarchically, with parallel exploration across branches and sequential refinement within them, as in Tree-of-Debate~\cite{164} and MAKGED~\cite{82}. 
Hybrid mechanisms trade off exploration and coordination --- parallel phases promote diversity and efficiency, sequential phases enable convergence --- at the cost of greater orchestration complexity.

\subsection{Format}

\begin{definition}{Format}{def:format}
The representation and structuring of the information exchanged between agents. It defines how arguments are expressed, transformed, and delivered during the debate.
\end{definition}

The format specifies how messages are represented and processed. For a message $m_{u \rightarrow v}^{(t)}$ transmitted along an edge $(u,v) \in E$ at step $t$, it determines the encoding used, whether and how the message is transformed before consumption, and how much prior context is available. % --- decisions that directly affect interpretability, information fidelity, and communication efficiency.

\subsubsection{Syntax}

Syntax is the representational form in which messages are encoded and exchanged. It governs how information is structured, interpreted, and processed across agents, affecting transparency, inter-agent compatibility, and how far intermediate reasoning can be inspected or manipulated.

A \emph{natural language} syntax encodes messages as human-readable text and is almost exclusively the choice in MAD (99.3\%). It enables rich, expressive, and interpretable exchanges --- allowing agents to articulate reasoning, critique others, and build on prior arguments --- and supports human-in-the-loop supervision, but introduces ambiguity, verbosity, and higher token consumption.

An \emph{embeddings}-based syntax exchanges messages as dense vectors in latent space rather than explicit text. The only approach proposing and evaluating it is CIPHER~\cite{17}, where agents generate messages as weighted averages of token embeddings drawn from the output distribution and exchange these directly, converting back to text only at the final stage via nearest-neighbour search; Pham et al. report consistent gains over natural-language debate (0.5--5.0\%), particularly for smaller models~\cite{17}. 
This can reduce communication overhead and enable compact exchange, but limits interpretability and fine-grained critique, as intermediate representations are not human-readable.

\subsubsection{Content}

Content refers to the degree of semantic manipulation applied to a message before transmission. Formally, the delivered content is $\tilde{m}_{u \rightarrow v}^{(t)} = f(m_{u \rightarrow v}^{(t)})$. Content choice directly affects the fidelity, conciseness, and informational density of the exchanged arguments.

A \emph{verbatim} strategy (86.1\%) transmits messages without modification ($\tilde{m}_{u \rightarrow v}^{(t)} = m_{u \rightarrow v}^{(t)}$), preserving the full original argument. The vast majority of approaches  exchange outputs as-is~\cite{1,12,25,28,29,31,36,39,40,51,82}. Verbatim exchange is often augmented with metadata that accompanies the argument without altering it:

\begin{itemize}
    \item \textbf{Confidence scores}, signalling an agent's degree of certainty and used to weight aggregation, as in RECONCILE's confidence-weighted voting~\cite{7,21,72}.
    \item \textbf{Uncertainty scores}, typically the inverse of confidence, letting agents adjust attention towards more reliable contributions~\cite{66}.
    \item \textbf{Verifier or evaluation scores}, where a dedicated component scores answers and intermediate reasoning; Park et al.~\cite{99} aggregate these into rewards optimized through multi-agent reinforcement learning.
    \item \textbf{Judge-based likelihoods}, where a judge assigns a plausibility signal used for final selection without altering the arguments themselves~\cite{124}.
    \item \textbf{Bias or quality scores}, computed by auxiliary agents to guide selection or training (reinforcing desirable behaviours, penalizing flawed outputs) while leaving arguments unchanged~\cite{133}.
    \item \textbf{Guideline or preference weights}, external signals encoding priorities such as utility or fairness that modulate aggregation without modifying the exchanged content~\cite{27}.
\end{itemize}

Verbatim exchange maximizes information fidelity and retains full reasoning traces for inspection and post-hoc analysis, but introduces redundancy and higher token cost.

A \emph{summarized} strategy (10.6\%) compresses the message, $\tilde{m}_{u \rightarrow v}^{(t)} = f(m_{u \rightarrow v}^{(t)})$ with $|f(m)| < |m|$, primarily to control context growth and communication cost~\cite{9,44,68,105}. Some approaches replace full dialogue history with compact shared summaries that preserve key signals~\cite{37,67}; others summarize more aggressively at the message level, via entropy-based compression of redundant content~\cite{105}, coordinator agents that synthesize outputs into concise reports as in ColaCare~\cite{44}, or judge agents that summarize each round before continuing~\cite{68}. Empirically, summarization can substantially cut communication overhead while remaining competitive~\cite{37,46,67}, and some works combine verbatim and summarized exchange by role~\cite{80,103,105}, preserving detail during local reasoning while compressing for coordination or aggregation. This improves efficiency and scalability but risks information loss when critical details are omitted.

\emph{Other} strategies transform the structure, selection, or informational properties of messages rather than simply transmitting (verbatim) or compressing (summarized) them. Estornell et al.~\cite{114} apply \emph{diversity-pruning} and \emph{quality-pruning} to selectively filter and prioritize responses, increasing diversity and reducing echo-chamber effects, while Wang et al.~\cite{118} reorganize intermediate outputs into staged reasoning pipelines, altering how information is decomposed and recombined. These reshape properties such as diversity or relevance and can mitigate failure modes like consensus collapse, at the cost of added design complexity and reduced transparency.

\subsubsection{Memory access}

Memory access refers to the extent and source of information available to an agent when generating and interpreting messages. For a message $m_{u \rightarrow v}^{(t)} = g(\mathcal{M}^{(t)})$, the accessible memory $\mathcal{M}^{(t)}$ may span prior debate messages, persistent knowledge, or external resources, directly affecting reasoning continuity, knowledge grounding, and coherence.

A \emph{short-term} strategy (94.7\%) restricts $\mathcal{M}^{(t)}$ to the current debate, typically only the messages exchanged within the ongoing interaction~\cite{16,20,29,50,58,59,63}. This is by far the most prevalent setting, usually implemented by concatenating prior arguments into the prompt at each step~\cite{16}, sometimes with a compression step to keep within model context limits~\cite{50} --- a concern that directly motivates summarization. It simplifies reproducibility and experimental control, since sessions are self-contained, but prevents agents from reusing knowledge accumulated across prior debates.

A \emph{long-term} strategy (2.7\%) extends $\mathcal{M}^{(t)}$ with information persisting across sessions, enabling cumulative reasoning, refinement, and adaptation over time~\cite{55,61,78,105}. It is typically implemented through persistent debate traces, structured memory modules, historical reflections, or error repositories revisited in future steps~\cite{55,78,105}, sometimes combined with learning-based adaptation that refines behaviour across iterations~\cite{55,61}. It raises challenges of memory management, trace selection, and consistency, since outdated, noisy, or erroneous information may persist.

A \emph{knowledge retrieval} strategy (6.0\%) augments $\mathcal{M}^{(t)}$ with external sources --- databases, documents, guidelines, or retrieval systems queried during the debate --- to compensate for insufficient task knowledge and improve factual grounding~\cite{8,10,23,36,44,70,75,78,164}. Examples include medical guidelines in clinical consultation~\cite{44}, stance-related background in zero-shot stance detection~\cite{75}, event-schema evidence in event extraction~\cite{10}, and paper-specific passages in scientific comparison~\cite{164}; others use shared or adaptive knowledge pools that agents query on demand~\cite{70}. This enhances grounding and domain coverage but makes debate quality dependent on retrieval accuracy, evidence selection, and latency.

Finally, a small subset (2.0\%) explores \emph{parametric memory}, embedding debate-derived knowledge directly into model parameters through learning rather than explicitly accessible context~\cite{55,61,105}. These approaches use debate traces, critiques, error logs, or reinforcement signals to adapt agent behaviour across training cycles: refining policies towards safer, less toxic generations~\cite{61}, leveraging traces as synthetic supervision for self-evolution~\cite{55}, or combining adversarial debate with persistent error logging for hallucination mitigation~\cite{105}. 
Parametric memory adds no prompt context and needs no retrieval at inference, making it efficient once trained, but introduces training complexity and risks such as policy drift or the absorption of low-quality debate-derived patterns.

\begin{rqfindings}{3}
\begin{itemize}
    \item[\faHandPointRight] \textbf{Dominance of full connectivity as a default design choice}. Most MAD approaches rely on fully connected topologies, favouring maximal information sharing over deliberate communication design. Yet structured alternatives such as sparse, ring, or group-divided topologies can achieve competitive performance while reducing token cost, leaving topology an underexploited design lever~\cite{2,3,22,81}.

    \item[\faHandPointRight] \textbf{Convergence on a static, verbatim, short-term interaction pattern}. The field has largely standardized around static structures, simultaneous or sequential natural-language exchange, verbatim message sharing, and short-term memory. This configuration is easy to implement, but only a few studies compare it against alternatives, leaving trade-offs in accuracy, cost, and communication efficiency unresolved~\cite{5,37,67}.

    \item[\faHandPointRight] \textbf{Neglect of ordering effects in sequential debate}. Sequential protocols are common, yet agent order is rarely controlled as an experimental variable. Since early contributions may anchor subsequent reasoning and amplify conformity, randomized turn-taking and centralized speaker selection remain important but underused mechanisms for reducing ordering bias~\cite{29,31,37,62}.

    \item[\faHandPointRight] \textbf{Limited adaptability due to session-scoped memory}. MAD systems almost universally rely on short-term memory restricted to the current debate, while long-term, retrieval-based, and parametric memory remain marginal. This limits agents' ability to accumulate knowledge across interactions or adapt to recurring tasks~\cite{55,61,105}.

    \item[\faHandPointRight] \textbf{Underexploration of non-verbal communication formats}. Inter-agent communication is almost exclusively natural language, despite early evidence that embedding-based communication can improve performance while reducing overhead. This suggests a largely unexplored space for efficient latent-space communication in larger or longer-running debate systems~\cite{17}.
\end{itemize}
\end{rqfindings}

\section{Agreement}
\label{sec:type-of-agreement}
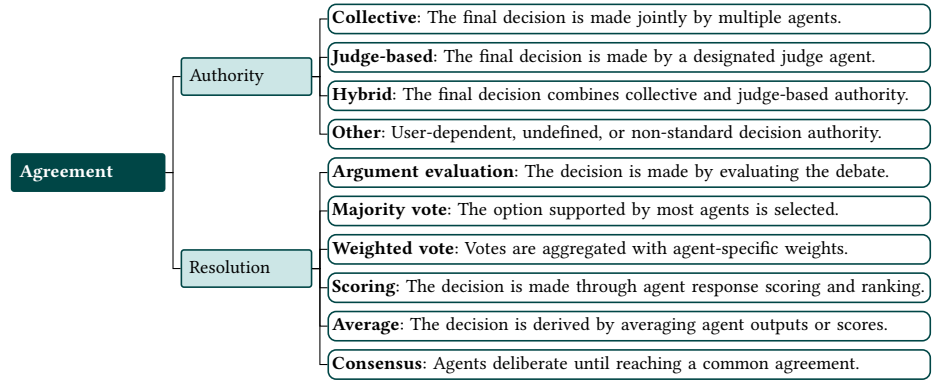
\begin{figure*}[ht]
    \centering
    \begin{forest}
    taxonomy tree
    [{\textbf{Agreement}}, rootbox
      [Authority, levelonebox
        [{\textbf{Collective}: The final decision is made jointly by multiple agents.}, refs],
        [{\textbf{Judge-based}: The final decision is made by a designated judge agent.}, refs],
        [{\textbf{Hybrid}: The final decision combines collective and judge-based authority.}, refs],
        [{\textbf{Other}: User-dependent, undefined, or non-standard decision authority.}, refs]
      ]
      [Resolution, levelonebox
        [{\textbf{Argument evaluation}: The decision is made by evaluating the debate.}, refs],
        [{\textbf{Majority vote}: The option supported by most agents is selected.}, refs],
        [{\textbf{Weighted vote}: Votes are aggregated with agent-specific weights.}, refs],
        [{\textbf{Scoring}: The decision is made through agent response scoring and ranking.}, refs],
        [{\textbf{Average}: The decision is derived by averaging agent outputs or scores.}, refs],
        [{\textbf{Consensus}: Agents deliberate until reaching a common agreement.}, refs]
      ]
    ]
    \end{forest}
    \caption{Taxonomy of agreement in multi-agent debate systems.}
    \Description{A taxonomy tree for the agreement dimension in multi-agent debate strategies. The root is Agreement. It has two branches: Authority and Resolution. Authority includes Collective, Judge-based, Hybrid and Other. Resolution includes Argument Evaluation, Majority vote,  Weighted vote, Scoring, Average, and Consensus.}
    \label{fig:mad-agreement-tree}
\end{figure*}

The following section examines how MAD frameworks conclude their interactions and reach a final output. 
This process is categorized into two main dimensions, summarized in the taxonomy shown in Figure~\ref{fig:mad-agreement-tree}: the \textit{Authority} (who makes the decision) and the \textit{Resolution} mechanism (how the decision is determined).

\subsection{Authority}

\begin{definition}{Authority}{def:authority}
Authority refers to the agent responsible for the final output after the agents have concluded their debate process.
\end{definition}

The authority defines the agency or entity responsible for making the final determination and generating the system's output after the agents have concluded their debate process. It distinguishes between decisions made jointly by the group (\emph{Collective}), by a specialized agent (\emph{Judge-based}), through a combination of both (\emph{Hybrid}), or other non-standard models (\emph{Other}).

\begin{itemize}
    \item \textbf{Collective} (\emph{52.3\%}): This is the most frequent approach, where the final answer is an aggregation of the agents' positions. It is highly prevalent in mathematical and reasoning tasks where objective answers are required~\cite{2, 17, 18, 24, 33, 37, 49, 128}. 

    \item \textbf{Judge-based} (\emph{39.1\%}): A designated agent --- often referred to as a \emph{``Judge'', ``Moderator'', ``Captain'', or ``Meta-Agent''} --- reviews the entire debate log to make the final determination~\cite{1, 3, 4, 8, 10, 11, 29, 31, 36, 40, 44, 48, 50, 56, 57, 101, 103, 106, 122, 126}. Although judges typically access the full conversation, some specialized designs only provide the judge with a summary generated by a separate agent~\cite{4}. While this model is typical for subjective tasks such as text evaluation or medical diagnostics~\cite{4, 56, 101}, it is also applied in scientific comparative analysis, where a moderator synthesizes the entire debate tree into a paragraph-long comparative summary~\cite{164}.  
    
    \item \textbf{Hybrid} (\emph{4.0\%}): These systems combine both methods, usually attempting a collective agreement first and invoking a judge only if a tie occurs or consensus is not reached within a specific number of rounds~\cite{22, 76, 82, 114, 121}.
    
    \item \textbf{Other} (\emph{4.6\%}): This category encompasses frameworks where the decision authority is not explicitly defined, is user-dependent, or follows non-standard protocols. It includes \emph{human-in-the-loop} systems where the final decision is delegated to the user~\cite{43}, and fully configurable platforms that allow for any agreement style depending on the setup~\cite{125}. Furthermore, it accounts for mathematical resolutions such as reaching a \emph{Nash equilibrium} to achieve cross-cultural consensus~\cite{27}, or architectural designs where a \emph{Central Answer Model (CAM)}, consisting of fine-tuned ML models, generates the final response~\cite{115}. Other proposals include selecting the output of a specific agent, like the \emph{``harmless agent''} after a debate~\cite{61}.

\end{itemize}

%In some approaches, agreement is reached \textit{collectively} by the debaters themselves. Although the goal of most MAD approaches is to reach consensus, this might not always be the case even after a long debate. In these cases, the final collective agreement is obtained by \textit{majority vote}~\cite{2,7,9,12,14,17,18,24}, \textit{weighted or scoring-based vote} based on the weighted votes obtained from the debaters~\cite{5,21}, or \textit{averaging} the responses if they are numerical~\cite{12,14}.

%Other family of approaches rely on a judge to determine the most convincing argument~\cite{1,3,4,8,15,16}. This \textit{judge-based agreement} is especially common again in bilateral structures. The judge is usually given access to the entire debate, but we found some approaches where only a summary provided by a summarizer was given~\cite{4}. In some cases, the judge’s role is limited to deciding when consensus has been reached, while the summarizer or editor produces the final answer~\cite{8,10}. 
%In one instance, we found an hybrid approach, where the judge is only used to decide if a tie has been reached after the votes of the debaters~\cite{22}.

\subsection{Resolution}

\begin{definition}{Resolution}{def:resolution}
The resolution mechanism specifies the logic or algorithm used to select or generate the final response.
\end{definition}

Resolution is the specific mechanism, logic, or algorithmic process employed to aggregate individual agent positions or synthesize the arguments exchanged during the debate into a single, unified final response. This includes methods such as majority voting, scoring, numerical averaging, or reaching a consensus.

\begin{itemize}
    \item \textbf{Argument Evaluation} (\emph{13.4\%}\footnote{Exclusively, for resolution, percentages are computed over $n=97$ MAD approaches, representing the set of studies where a resolution agreement process is actually formalized and implemented.}): Specific to judge-based systems, where the designated judge or meta-agent evaluates the quality of the debate arguments to synthesize a final judgment or summary~\cite{25, 52, 95, 101, 103, 111, 124, 127, 131, 136, 164}. This mechanism is essential for tasks where a binary choice is insufficient and a qualitative synthesis is required. For instance, in scientific comparative analysis, a moderator evaluates a debate tree to produce a paragraph-long summary highlighting significance and novelty~\cite{164}. Similarly, in software security, an ``Auditor'' reviews revised analyses to make final determinations on vulnerability classifications~\cite{95}.

    \item \textbf{Majority vote} (\emph{46.4\%}): The dominant mechanism in collective systems, where the answer selected by the highest number of agents is chosen as the final outcome~\cite{2, 9, 12, 17, 18, 24, 33, 37, 49, 55, 60, 62, 63, 66, 68, 72, 74, 78, 81, 91, 99, 105, 118, 128}. A notable special case occurs in frameworks like \textit{Auto-Arena}, where although the authority is judge-based, the final decision is delegated to a \emph{committee of judges} who reach a verdict through a majority vote among themselves~\cite{35}.

    \item \textbf{Weighted vote} (\emph{5.2\%}): In this mechanism, final decisions are reached by applying specific weights to agent responses rather than treating them equally~\cite{21, 27, 45, 70, 167}. Weights are typically calculated based on metadata such as confidence levels~\cite{45} or an agent's consistency across multiple rounds of debate~\cite{45}. These weights are processed through quantitative mathematical formulas --- such as specific equations to calculate final scores~\cite{45} --- or derived from \emph{Nash equilibrium} distributions in cross-cultural negotiation tasks~\cite{27}. Other implementations use algorithmic weighted predictions to combine adversarial perspectives into a single output~\cite{70}.

   \item \textbf{Scoring} (\emph{10.3\%}): In this mechanism, agents assign numerical values or ranks to different responses, and the system selects the option with the best aggregate performance. This is used in spam detection to classify emails~\cite{5} and in math word problem solving to identify the most plausible reasoning path~\cite{23, 26}. A sophisticated application is found in \emph{value alignment} tasks, where responses are selected by maximizing a ``usefulness'' function while simultaneously minimizing ``harmlessness'' scores~\cite{80}. Additionally, scoring can be used by committees of judges to detect misinformation~\cite{58} or to categorize arguments based on their bias levels~\cite{133}.
   
   \item \textbf{Average} (\emph{4.1\%}): This method is specifically employed when the agents' outputs are numerical. Instead of choosing one agent's response, the system computes the mathematical mean of all individual values to derive a unified final answer. This is a common practice in evaluation frameworks like ChatEval~\cite{12} and in studies on confidence calibration, where numerical scores are averaged to provide a more reliable assessment of truthfulness~\cite{14}.

    %\item \textbf{Scoring and Averaging:} Agents assign numerical scores or values to the proposed outputs. The final decision is determined by ranking (selecting the response that maximizes a specific metric) or by computing the mathematical average of all the numerical responses of the agent~\cite{5, 12, 14, 23, 26, 80, 133}.
    
    \item \textbf{Consensus} (\emph{20.6\%}): Agents deliberate until they all converge on a single, unanimous answer~\cite{19, 46, 51, 54, 59, 64, 67, 93, 96, 98, 123}. To prevent infinite loops, these processes are usually capped at a maximum number of rounds. For example, Chen et al. assessed the impact of doubling debate rounds (e.g., from 2 to 4, from 13 to 25) observing performance and token efficiency~\cite{64}.  
\end{itemize}

%Together, these dimensions illustrate the richness and diversity of methodologies in MAD approaches, highlighting both the flexibility of the paradigm and the patterns that shape its current landscape.

\begin{rqfindings}{4}
\begin{itemize}
    \item[\faHandPointRight]\textbf{Dominance of collective authority in objective domains}. Research shows a clear preference for collective decision-making in domains with objective ground truths, such as mathematics and logic~\cite{2, 17, 18, 24, 49, 128}. In these contexts, authority is shared among agents to ensure that the final output is a product of group convergence.

    \item[\faHandPointRight] \textbf{Specialization of judge-based authority (Meta-Agents) for qualitative synthesis}. There is a significant trend towards delegating authority to specialized ``Meta-Agents'' or judges for tasks requiring nuanced synthesis, such as clinical diagnostics or open-ended text evaluation~\cite{1, 3, 35, 101, 103, 164}. These judges move beyond simple aggregation to perform a qualitative evaluation of the debate's arguments.

    \item[\faHandPointRight]\textbf{Strategic trade-off between resolution efficiency and output fidelity}. The choice of resolution mechanism involves a trade-off: while majority voting is the standard for computational efficiency in objective tasks~\cite{2, 49, 72}, mechanisms like consensus or iterative argument evaluation prioritize high-fidelity results at a higher cost in terms of communication rounds and token consumption~\cite{54, 64}.
\end{itemize}
\end{rqfindings}

\section{Discussion}
\label{sec:discussion}

\textbf{Documenting the MAD research landscape.}
The first contribution of this study is to serve as a structured proxy for the state of MAD research. The bibliometric trend is unambiguous: from 14 studies in 2023 to 70 in 2025, with 45 of 141 studies available only as preprints, MAD is a young, fast-moving field whose body of knowledge remains largely uncurated and fragmented across domains and terminologies. 
Reading the four research questions jointly reveals how design decisions are coupled. 
Domain choice (RQ1) propagates into participant design (RQ2): objective-ground-truth domains such as mathematics and coding favour homogeneous debaters converging on a single answer, whereas less objective domains such as healthcare or social reasoning motivate heterogeneous personas and explicit roles to broaden perspective coverage. 
The same domains shape agreement (RQ4): objective tasks gravitate towards collective authority resolved by majority vote, while subjective tasks delegate to judge-based authority and argument evaluation, where a meta-agent performs qualitative synthesis rather than aggregation. Interaction (RQ3) shows the strongest internal coupling: across topology, protocol, format, and memory, the field has implicitly converged on a recurring pattern --- static, fully connected topologies, verbatim natural-language content, and short-term memory --- adopted by convention rather than through systematic comparison. 
Crucially, the alternatives to this pattern are not absent but marginal: dynamic topologies, structured and sparse networks, summarized or embedding-based content, and long-term, retrieval-augmented, or parametric memory each appear in a small minority of studies, and the few works that compare them report non-trivial performance and efficiency gains. 
\textbf{The taxonomy presented in this paper is the first-level contribution that makes these observations possible}. 
Decomposing MAD into \emph{participants}, \emph{interaction}, and \emph{agreement}, and each dimension into orthogonal sub-dimensions with explicit attribute values, provides a shared vocabulary that replaces the inconsistent and overloaded terminology currently found in the literature. %, acting simultaneously as a map of what has been done and a checklist of what remains open.

This consolidated view exposes a set of concrete and actionable research gaps. Each identifies a design dimension that is technically feasible to vary but has not been investigated in a controlled manner:

\begin{itemize}
    \item \textbf{Underexplored domains.} Application is concentrated on general reasoning and mathematics, while specialized professional domains remain comparatively untested.
    \item \textbf{Topology as a deliberate design lever.} Full connectivity is adopted by default, yet sparse and structured alternatives match or exceed it at lower cost. The conditions under which a given topology is preferable remain uncharacterized.
    \item \textbf{Dynamic interaction.} Adaptive topologies are rare despite their potential to prune redundant exchanges. Principled policies for when and how to evolve the communication graph are largely missing.
    \item \textbf{Interaction order.} Agent ordering in sequential protocols is seldom controlled or reported, leaving an underappreciated source of anchoring and conformity bias.
    \item \textbf{Memory beyond the session.} Long-term, retrieval-augmented, and parametric memory are explored only marginally, limiting the adaptability of MAD systems across recurring tasks.
    \item \textbf{Embedding-based communication.} Latent-space message exchange is virtually unexplored despite early evidence of joint effectiveness and efficiency gains.
    \item \textbf{Cross-dimension interactions.} Studies typically vary one dimension in isolation; how participant, interaction, and agreement choices interact --- e.g., topology effects under different agreement mechanisms --- is unstudied.
    \item \textbf{Cost-aware evaluation.} Performance is rarely reported jointly with token consumption or latency, obscuring the efficiency trade-offs that several design choices are explicitly meant to address.
\end{itemize}

\textbf{A structural framework for benchmarking.}
A direct practical consequence of the taxonomy is that it reframes the comparison of MAD approaches. 
As the results show, a MAD setting is defined by a large number of design decisions --- on the order of a dozen attributes, each admitting several values --- which jointly induce an enormous configuration space. 
This combinatorial variance is the root cause of a recurring methodological weakness: comparisons across studies are difficult, and rigorous benchmarking is harder still, because reported results conflate the effect of the debate mechanism with the effect of unstated design choices. 
The taxonomy mitigates this by giving comparison a format. Instead of relying on qualitative, narrative descriptions, an approach can be specified as a point in the taxonomic space, making explicit which attributes are fixed and which are varied. This yields two concrete benefits. 
First, \emph{differential documentation}: it becomes straightforward to express how one MAD setting extends or modifies another by reporting only the attributes that differ, so that ablations and incremental contributions are described precisely instead of in prose. 
Second, \emph{controlled variability}: reported performance can be attributed to specific attributes, enabling studies that isolate the effect of a single design dimension while holding the rest constant --- the prerequisite for fair benchmarking and for distinguishing genuine mechanism gains from configuration artifacts. 
In this sense, the taxonomy is not only descriptive but normative, defining the minimal set of dimensions a study must report for its results to be interpretable and reproducible.

\textbf{Towards machine-readable MAD specifications.}
The third perspective concerns design and implementation. Because the taxonomy is structured and its attributes take values from well-defined domains, it can be extended beyond documentation into a machine-readable specification of MAD systems. 
Several dimensions already admit formal notations that make this immediate: topology is naturally expressed as a graph $G=(V,E)$, the protocol as a temporal activation of edges $E(t)$, the format as transformation functions over messages, and agreement as an authority-and-resolution pair. These are constructive devices, in the sense that a configuration expressed in this notation is sufficient to instantiate the corresponding system. This opens a path towards treating MAD settings as \emph{artifacts}: structured, declarative specifications that can be exchanged, version-controlled, and modified independently of any particular implementation. 
Two consequences follow. First, reproducibility improves, since a specification artifact removes ambiguity about the configuration under evaluation. 
Second, and more importantly, the configuration space becomes programmatically navigable: once a MAD setting is a structured object, attributes such as the number of debaters, topology, or maximum rounds become hyperparameters amenable to automated search and tuning, rather than choices fixed manually at design time. The taxonomy thus provides the schema on top of which automated MAD design and optimization pipelines can be built.

\begin{table*}[ht]
\centering
\caption{Taxonomy instantiation of ChatEval~\cite{12}, using the semi-formal notation introduced in Sections~\ref{sec:participants}--\ref{sec:type-of-agreement}.}
\label{tab:chateval-instantiation}
\renewcommand{\arraystretch}{1.4}
\setlength{\tabcolsep}{4pt}
\begin{tabular}{@{} l l l p{3.0cm} p{3.2cm} p{3.2cm} @{}}
\toprule
\textbf{Dim.} & \textbf{Sub-dim.} & \textbf{Attribute} &
  \textbf{S1: One-by-One} &
  \textbf{S2: Simultaneous-Talk} &
  \textbf{S3: Simult.-Talk + Summarizer} \\
\midrule

\multirow{6}{*}{\rotatebox[origin=c]{90}{\textsc{Participants}}}
  & Base Model & Family
    & \multicolumn{3}{c}{Closed-weight (GPT-4 / GPT-3.5-turbo)} \\
\cmidrule(l){2-6}
  & Base Model & Heterogeneity
    & \multicolumn{3}{c}{Homogeneous: $\forall u,v \in V,\; \text{model}(u) = \text{model}(v)$} \\
\cmidrule(l){2-6}
 & Roles & ---
    & \multicolumn{2}{p{6.4cm}}{$V = V_D = \{$Alice, Bob, Carol$\}$ = Debaters}
    & $V = V_D \cup \{S\}$ where $V_D$ = Debaters and $S$ = Summarizer \\
    \cmidrule(l){2-6}
  & Personas & Attributes
    & \multicolumn{3}{c}{Different personality traits through role prompts} \\
\cmidrule(l){2-6}
  & Personas & Heterogeneity
    & \multicolumn{3}{p{9.6cm}}{Heterogeneous (default): distinct personality traits per agent; homogeneous baseline also evaluated (same prompt for all $u \in V_D$)} \\
\cmidrule(l){2-6}
  & Personas & Assignment
    & \multicolumn{3}{c}{Manually identified at design time} \\
\midrule
\multirow{9}{*}{\rotatebox[origin=c]{90}{\textsc{Interaction}}}
  & Topology & Adaptability
    & \multicolumn{3}{c}{Static: $G = (V, E)$ invariant across all rounds $t \in T$} \\
\cmidrule(l){2-6}
  & Topology & Structure
    & \multicolumn{2}{p{6.4cm}}{Fully connected over $V_D$:
      $E = \{(u,v) \mid u,v \in V_D,\; u \neq v\}$, $|E|=6$}
    & Fully connected over $V_D$ plus unidirectional edges from each debater to Summarizer:
      $E = E_D \cup \{(d, S) \mid d \in V_D\}$, $|E_D|=6$, $|E|=9$ \\
\cmidrule(l){2-6}
  & Protocol & Mechanism
    & Sequential: $|E(t)|=1$; agents traverse $V_D$ in fixed order Alice $\to$ Bob $\to$ Carol per round $t$; prior responses concatenated into $H_n$
    & Simultaneous: $|E(t)|=|E_D|$; all $u \in V_D$ generate $m_{u\to *}^{(t)}$ in parallel; outputs buffered and broadcast to all $H_n$ after each round
    & Simultaneous: same as S2 for $V_D$; Summarizer $S$ additionally activated once per round after buffer is complete \\
\cmidrule(l){2-6}
  & Format & Syntax
    & \multicolumn{3}{c}{Natural language: $m_{u\to v}^{(t)} \in \Sigma^*$ for all $(u,v) \in E$} \\
\cmidrule(l){2-6}
  & Format & Content
    & Verbatim: $\tilde{m}_{u\to v}^{(t)} = m_{u\to v}^{(t)}$; full prior history $H_n$ concatenated before each turn
    & Verbatim: $\tilde{m}_{u\to v}^{(t)} = m_{u\to v}^{(t)}$; round outputs buffered and appended to all $H_n$ simultaneously
    & Hybrid: verbatim among debaters, $(u,v) \in E_D$: $\tilde{m}_{u\to v}^{(t)} = m_{u\to v}^{(t)}$; summarized from $S$ to $V_D$: $\tilde{m}_{S\to d}^{(t)} = f(\text{buf}^{(t)})$% where $f = \text{SUM}(\cdot)$ and $|\tilde{m}_{S\to d}^{(t)}| < |\text{buf}^{(t)}|$ 
    \\
\cmidrule(l){2-6}
  & Format & Memory $\mathcal{M}^{(t)}$
    & \multicolumn{3}{p{9.6cm}}{Short-term: $\mathcal{M}^{(t)} = \{m_{u\to v}^{(t')} \mid t' < t,\; (u,v) \in E\}$; session-scoped} \\
\midrule
\multirow{2}{*}{\rotatebox[origin=c]{90}{\textsc{Agrmt.}}}
  & Authority & ---
    & \multicolumn{3}{c}{Collective: final decision made jointly by all debaters in $V_D$} \\
\cmidrule(l){2-6}
  & Resolution & ---
    & \multicolumn{3}{p{9.6cm}}{Hybrid: majority vote if output is categorical; average if output is numerical} \\
\bottomrule
\end{tabular}
%\smallskip
%\noindent\footnotesize
%\textit{Notation}: $V$ = full agent set; $V_D \subseteq V$ = debater subset; $S \in V \setminus V_D$ = summarizer agent; $E$ = full edge set; $E_D = \{(u,v) \mid u,v \in V_D, u \neq v\}$ = debater-only edges; $E(t) \subseteq E$ = active edges at time step $t$; $m_{u\to v}^{(t)}$ = message transmitted from agent $u$ to agent $v$ at step $t$; $\tilde{m}$ = delivered (possibly transformed) message; $f$ = content transformation function; $\text{buf}^{(t)}$ = buffer of all debater outputs at round $t$; $\mathcal{M}^{(t)}$ = memory context available at step $t$; $H_n$ = chat history of agent $n$; $\Sigma^*$ = set of natural language token sequences.
\end{table*}

\textbf{A formalization example: ChatEval.}
To illustrate the three perspectives concretely, Table~\ref{tab:chateval-instantiation} applies the taxonomy to ChatEval~\cite{12}, using the formal notation introduced in \Cref{sec:participants,sec:interaction,sec:type-of-agreement}. We select ChatEval as it is among the most highly cited frameworks in our corpus, exercises all three taxonomy dimensions non-trivially, and is representative of the evaluation-oriented MAD systems common in the literature. 
The exercise instantiates each dimension --- participants, interaction, and agreement --- for the three MAD strategies that ChatEval defines, and exemplifies the benefits discussed above.
As \emph{documentation}, it renders the complete configuration of a published approach explicit and unambiguous. 
As a \emph{benchmarking} aid, it shows that the three strategies differ in only a small set of attributes --- the protocol mechanism, and the presence of a summarizer role with the associated topology and content changes --- while all remaining attributes are shared, making the comparison precise and the contribution of each variant isolable. 
As a \emph{specification}, the formal entries ($G=(V,E)$, $E(t)$, $\tilde{m}=f(m)$, authority--resolution pair) constitute a near-constructive description from which the system could be instantiated. This single example thus demonstrates how the taxonomy operates simultaneously as a descriptive, comparative, and constructive instrument.

\section{Threats to Validity}
\label{sec:threats-to-validity}

We organize threats to validity following the taxonomy proposed by Zhou et al.~\cite{Zhou2016}.

Concerning \textit{internal validity}, two threats arise from the human-dependent nature of the study. First, paper selection and annotation are susceptible to subjectivity. To mitigate this threat, every candidate paper was independently assessed by at least two authors, with disagreements resolved by consensus. 
Data extraction was performed by a single author per paper due to corpus size, but the taxonomy vocabulary was iteratively consolidated across all authors throughout the process. 
Second, the taxonomy was derived inductively from the literature rather than from a pre-established framework, meaning its categories may be shaped by the particular set of studies included. To mitigate this threat, we iteratively reviewed and expanded the vocabulary included in our taxonomy, revisiting studies as new paradigms or design alternatives emerged.

Concerning \textit{construct validity}, the main threat is the absence of a universally accepted definition of MAD. We addressed this by deriving explicit inclusion and exclusion criteria (Table~\ref{tab:criteria}) from a synthesis of existing definitions. 
Nevertheless, the boundary with adjacent paradigms such as multi-agent collaboration or iterative self-refinement is inherently fuzzy, and borderline cases required judgment calls that the dual-review protocol only partially mitigates. 

Concerning \textit{external validity}, the study is scoped to natural-language tasks, excluding multimodal debate settings. This is a deliberate decision to ensure taxonomic coherence, and we identify multimodal MAD as a direction for future work. 
Consequently, our taxonomy generalizes to text-based MAD and may need to be extended to capture modality-specific dimensions. Nevertheless, the majority of design dimensions are independent from the modality of the debate.

Concerning \textit{conclusion validity}, the main threat concerns the completeness and repeatability of the search. The primary search was conducted exclusively on Scopus using a deliberately narrow search string, prioritizing precision in the seed set at the cost of potentially missing relevant works. This is partially mitigated by the systematic snowballing process, which traces backward and forward citation links to recover relevant works beyond the initial query, including arXiv preprints. 
Snowballing was executed in a single iteration rather than until full saturation; given that 70 of the 141 included studies were published in 2025 alone, additional rounds would yield diminishing returns, as later-included papers would largely cite works already in the corpus. Finally, the rapid evolution of the field means that findings and taxonomic categories may require revision as new architectural patterns emerge. However, this mostly affects the coverage and distribution of surveyed studies, rather than the validity of the design dimensions covered by our taxonomy and the dimension-dependent insights derived from the set of surveyed studies.
To support external auditability and incremental updates, we make our full replication package available~\cite{Motger2026MADReplication}.

\section{Conclusions}
\label{sec:conclusions}

This paper surveyed MAD research through a systematic review of 141 primary studies published between 2023 and 2025. From this corpus, we derived a three-dimensional taxonomy --- \emph{participants}, \emph{interaction}, and \emph{agreement} --- that consolidates a fragmented and terminologically inconsistent body of work into a shared, formally grounded vocabulary.

Two observations stand out. 
First, the field has converged on a narrow design pattern --- e.g., static, fully connected topologies with simultaneous protocols, verbatim natural-language exchange, and short-term memory. Design alternatives such as sparse topologies, embedding-based communication, and persistent memory remain marginal, despite early evidence in their favour. 
Second, any MAD setting is the product of roughly a dozen interacting design decisions, which makes cross-study comparison unreliable whenever those decisions are left implicit.
Together, these observations point to the same underlying gap: MAD design choices are made by convention and reported in prose, leaving the paradigm without a principled basis for knowing which configurations work, when, and why.

Our taxonomy responds to both. It serves as a map of the current research landscape and a checklist of underexplored directions, but its more distinctive role is structural. As most of its dimensions admit formal notation, several design aspects of any MAD configuration can be expressed as an explicit, machine-readable artifact rather than a prose description. Treating debate settings as artifacts turns the configuration space into an object that can be documented, compared attribute by attribute, version-controlled, and ultimately searched automatically.

We see this as the principal opportunity ahead. The immediate value of the taxonomy is descriptive, but its longer-term value is operational --- i.e., a schema on which benchmarking protocols and automated configuration tooling can be built. As future work, we plan to formalize this schema into an executable specification and to use it for controlled, cost-aware benchmarking that isolates the effect of individual design dimensions. Extending the scope to multimodal debate remains a further open direction. 

\section*{Acknowledgments}
This work has been supported by funding from the HIVEMIND project – Horizon Europe call HORIZON-CL4-2024-DIGITAL-EMERGING-01 under Grant Agreement Number 101189745.

\section*{Data Availability Statement}
\label{sec:das}

A replication package containing the full list of surveyed studies, the data extraction sheets, and the taxonomy coding schema is publicly available on \href{https://github.com/nlp4se/MAD-rep-package}{GitHub} and archived on Zenodo with a permanent DOI~\cite{Motger2026MADReplication}.

\section*{Use of Generative AI Statement}

Generative AI tools were used exclusively to support language refinement and polishing of the manuscript. These tools were not used for literature analysis, data extraction, interpretation of results, or generation of scientific claims. All scientific content, methodological decisions, analyses, and conclusions were developed and validated by the authors.

% \section*{CRediT authorship contribution statement}

% Quim Motger: Conceptualization, Methodology, Validation, Formal analysis, Investigation, Data curation, Project administration, Writing – original draft, Writing – review \& editing, Visualization.
% Marc Oriol: Conceptualization, Methodology, Validation, Formal analysis, Investigation, Data curation, Project administration, Writing – original draft, Writing – review \& editing, Visualization.
% Jordi Marco: Conceptualization, Methodology, Validation, Formal analysis, Investigation, Data curation, Writing – original draft, Writing – review \& editing, Visualization.
% Xavier Franch: Conceptualization, Methodology, Writing – review \& editing.

\bibliographystyle{ACM-Reference-Format}
\bibliography{references}

@article{Xi2025,
  author    = {Xi, Zhiheng and Chen, Wenxiang and Guo, Xin and others},
  title     = {The rise and potential of large language model based agents: a survey},
  journal   = {Science China Information Sciences},
  volume    = {68},
  number    = {2},
  pages     = {121101},
  year      = {2025},
  doi       = {10.1007/s11432-024-4222-0},
  issn      = {1869-1919}
}

@inproceedings{Yao2023ReAct,
  title     = {ReAct: Synergizing Reasoning and Acting in Language Models},
  author    = {Yao, Shunyu and Zhao, Jeffrey and Yu, Dian and others},
  booktitle = {Procs. of the 11th International Conference on Learning Representations},
  year      = {2023},
  url       = {https://arxiv.org/abs/2210.03629}}

@inproceedings{park2023generative,
author = {Park, Joon Sung and O'Brien, Joseph and Cai, Carrie Jun and others},
title = {Generative Agents: Interactive Simulacra of Human Behavior},
year = {2023},
isbn = {9798400701320},
doi = {10.1145/3586183.3606763},
booktitle = {Procs. of the 36th Annual ACM Symposium on User Interface Software and Technology},
articleno = {2},
numpages = {22},
}

@inproceedings{Wei2022CoT,
author = {Wei, Jason and Wang, Xuezhi and Schuurmans, Dale and others},
title = {Chain-of-thought prompting elicits reasoning in large language models},
year = {2022},
isbn = {9781713871088},
publisher = {Curran Associates Inc.},
booktitle = {Procs. of the 36th International Conference on Neural Information Processing Systems},
articleno = {1800},
numpages = {14},
location = {New Orleans, LA, USA}
}

@inproceedings{Wang2023SelfConsistency,
  title     = {Self-Consistency Improves Chain of Thought Reasoning in Language Models},
  author    = {Wang, Xuezhi and Wei, Jason and Schuurmans, Dale and Le, Quoc V. and Chi, Ed H. and Narang, Sharan and Chowdhery, Aakanksha and Zhou, Denny},
  booktitle = {Procs. of the 11th International Conference on Learning Representations},
  url = "https://arxiv.org/abs/2203.11171",
  year      = {2023}
}

@inproceedings{Yao2023TreeOfThought,
author = {Yao, Shunyu and Yu, Dian and Zhao, Jeffrey and others},
title = {Tree of thoughts: deliberate problem solving with large language models},
year = {2023},
booktitle = {Procs. of the 37th International Conference on Neural Information Processing Systems},
articleno = {517},
numpages = {14}
}

@inproceedings{Ouyang2023RLHF,
author = {Ouyang, Long and Wu, Jeff and Jiang, Xu and others},
title = {Training language models to follow instructions with human feedback},
year = {2022},
isbn = {9781713871088},
publisher = {Curran Associates Inc.},
booktitle = {Procs. of the 36th International Conference on Neural Information Processing Systems},
articleno = {2011},
numpages = {15}
}

@inproceedings{Rafailov2023DPO,
author = {Rafailov, Rafael and Sharma, Archit and Mitchell, Eric and others},
title = {Direct preference optimization: your language model is secretly a reward model},
year = {2023},
booktitle = {Procs. of the 37th International Conference on Neural Information Processing Systems},
articleno = {2338},
numpages = {14}
}

@inproceedings{Dettmers2023QLoRA,
author = {Dettmers, Tim and Pagnoni, Artidoro and Holtzman, Ari and Zettlemoyer, Luke},
title = {{QLORA}: efficient finetuning of quantized LLMs},
year = {2023},
booktitle = {Procs. of the 37th International Conference on Neural Information Processing Systems},
articleno = {441},
numpages = {28}
}

@inproceedings{Fan2024,
author = {Fan, Wenqi and Ding, Yujuan and Ning, Liangbo and others},
title = {A Survey on {RAG} Meeting {LLM}s: Towards Retrieval-Augmented Large Language Models},
year = {2024},
doi = {10.1145/3637528.3671470},
booktitle = {Procs. of the 30th ACM SIGKDD Conference on Knowledge Discovery and Data Mining},
numpages = {11}
}

@inproceedings{Hong2024MetaGPT,
  title     = {MetaGPT: Meta Programming for a Multi-Agent Collaborative Framework},
  author    = {Hong, Sirui and Zhuge, Mingchen and Chen, Jiaqi and others},
  booktitle = {Procs. of the 12th International Conference on Learning Representations},
  year      = {2024},
  url = "https://arxiv.org/abs/2308.00352"
}

@inproceedings{li-etal-2023-shot,
    title = "Few-shot In-context Learning on Knowledge Base Question Answering",
    author = "Li, Tianle  and
      Ma, Xueguang  and
      Zhuang, Alex  and
      Gu, Yu  and
      Su, Yu  and
      Chen, Wenhu",
    booktitle = "Procs. of the 61st Annual Meeting of the Association for Computational Linguistics (Volume 1: Long Papers)",
    year = "2023",
    doi = "10.18653/v1/2023.acl-long.385",
}

@misc{cheng2024adaptinglargelanguagemodels,
      title={Adapting Large Language Models to Domains via Reading Comprehension}, 
      author={Daixuan Cheng and Shaohan Huang and Furu Wei},
      year={2024},
      eprint={2309.09530},
      archivePrefix={arXiv},
      primaryClass={cs.CL},
}

@INPROCEEDINGS{Oriol2025,
  author={Oriol, Marc and Motger, Quim and Marco, Jordi and Franch, Xavier},
  booktitle={Procs. of the IEEE 33rd International Requirements Engineering Conference}, 
  title={Multi-Agent Debate Strategies to Enhance Requirements Engineering with Large Language Models}, 
  year={2025},
  volume={},
  number={},
  doi={10.1109/RE63999.2025.00063}}

@article{Wang2024,
  author    = {Wang, Lei and Ma, Chen and Feng, Xueyang and others},
  title     = {A Survey on Large Language Model Based Autonomous Agents},
  journal   = {Frontiers of Computer Science},
  year      = {2024},
  volume    = {18},
  number    = {6},
  pages     = {186345},
  doi       = {10.1007/s11704-024-40231-1},
  url       = {https://doi.org/10.1007/s11704-024-40231-1},
  issn      = {2095-2236}
}

@Article{Martín-Martín2021,
author={Mart{\'i}n-Mart{\'i}n, Alberto
and Thelwall, Mike
and Orduna-Malea, Enrique
and Delgado L{\'o}pez-C{\'o}zar, Emilio},
title={{G}oogle {S}cholar, {M}icrosoft {A}cademic, {S}copus, {D}imensions, {W}eb of {S}cience, and {O}pen{C}itations' {COCI}: a multidisciplinary comparison of coverage via citations},
journal={Scientometrics},
year={2021},
volume={126},
number={1}
}

@inproceedings{Wohlin2014,
author = {Wohlin, Claes},
title = {Guidelines for snowballing in systematic literature studies and a replication in software engineering},
year = {2014},
booktitle = {Procs. of the 18th International Conference on Evaluation and Assessment in Software Engineering},
articleno = {38},
numpages = {10},
series = {EASE '14}
}

@ARTICLE{Wang2026,
  author={Wang, Yuntao and Pan, Yanghe and Su, Zhou and Deng, Yi and Zhao, Quan and Du, Linkang and Luan, Tom H. and Kang, Jiawen and Niyato, Dusit},
  journal={IEEE Communications Surveys \& Tutorials}, 
  title={Large Model-Based Agents: State-of-the-Art, Cooperation Paradigms, Security and Privacy, and Future Trends}, 
  year={2026},
  volume={28},
  number={},
  pages={1906-1949},
  doi={10.1109/COMST.2025.3576176}}

@misc{Tillmann2025,
      title={Literature Review Of Multi-Agent Debate For Problem-Solving}, 
      author={Arne Tillmann},
      year={2025},
  archivePrefix = {arXiv},
  eprint = {2506.00066},
  journal = {arXiv}
}

@misc{SIGSOFT,
  author       = {{ACM SIGSOFT Empirical Standards}},
  title        = {Systematic Reviews Standard},
  howpublished = {\url{https://www2.sigsoft.org/EmpiricalStandards/docs/standards?standard=SystematicReviews\#}},
  note         = {Accessed: 2026-02-27},
  year         = {n.d.},
  organization = {ACM SIGSOFT Empirical Standards}
}

@article{Li2024,
  author    = {Li, Xinyi and Wang, Sai and Zeng, Siqi and Wu, Yu and Yang, Yi},
  title     = {A survey on LLM-based multi-agent systems: workflow, infrastructure, and challenges},
  journal   = {Vicinagearth},
  year      = {2024},
  volume    = {1},
  number    = {1},
  pages     = {9},
  doi       = {10.1007/s44336-024-00009-2},
  url       = {https://doi.org/10.1007/s44336-024-00009-2},
  issn      = {3005-060X},
  day       = {8}
}

@misc{Tran2025,
      title={Multi-Agent Collaboration Mechanisms: A Survey of {LLMs}}, 
      author={Khanh-Tung Tran and Dung Dao and Minh-Duong Nguyen and others},
      year={2025},
  archivePrefix = {arXiv},
  eprint = {2501.06322},
  journal = {arXiv}
}

@misc{Shen2023,
      title={Large Language Model Alignment: A Survey}, 
      author={Tianhao Shen and Renren Jin and Yufei Huang and Chuang Liu and Weilong Dong and Zishan Guo and Xinwei Wu and Yan Liu and Deyi Xiong},
      year={2023},
      eprint={2309.15025},
      archivePrefix={arXiv},
      primaryClass={cs.CL}
}

@article{Guan2026,
      title={Evaluating {LLM}-based Agents for Multi-Turn Conversations: A Survey}, 
      author={Shengyue Guan and Jindong Wang and Jiang Bian and others},
year = {2026},
volume = {17},
number = {4},
journal = {ACM Transactions on Intelligent Systems and Technology},
}

@article{Luo2025,
      title={Large Language Model Agent: A Survey on Methodology, Applications and Challenges}, 
      author={Junyu Luo and Weizhi Zhang and Ye Yuan and others},
      year={2025},
  archivePrefix = {arXiv},
  eprint = {2503.21460},
  journal = {arXiv}
}

@article{Li2024a,
      title={{LLM}s-as-{J}udges: A Comprehensive Survey on LLM-based Evaluation Methods}, 
      author={Haitao Li and Qian Dong and Junjie Chen and others},
      year={2024},
  archivePrefix = {arXiv},
  eprint = {2412.05579},
  journal = {arXiv}
}

@book{diestel2025graph,
  author    = {Diestel, Reinhard},
  title     = {Graph Theory},
  edition   = {6},
  series    = {Graduate Texts in Mathematics},
  volume    = {173},
  publisher = {Springer-Verlag},
  year      = {2025},
  isbn      = {978-3-662-70106-5},
  ean       = {978-3-662-70107-2},
  note      = {Sixth edition}
}

@inproceedings{1,
    title = "Counterfactual Debating with Preset Stances for Hallucination Elimination of {LLM}s",
    author = "Fang, Yi  and
      Li, Moxin  and
      Wang, Wenjie  and
      Hui, Lin  and
      Feng, Fuli",
    booktitle = "Procs. of the 31st International Conference on Computational Linguistics",
    year = "2025",
}

@inproceedings{2,
    title = "Improving Multi-Agent Debate with Sparse Communication Topology",
    author = "Li, Yunxuan  and
      Du, Yibing  and
      Zhang, Jiageng  and
      others",
    booktitle = "Findings of the Association for Computational Linguistics: EMNLP 2024",
    year = "2024",
    pages = "7281--7294"
}

@inproceedings{3,
    title = "Encouraging Divergent Thinking in Large Language Models through Multi-Agent Debate",
    author = "Liang, Tian  and
      He, Zhiwei  and
      Jiao, Wenxiang  and
      others",
    booktitle = "Procs. of the 2024 Conference on Empirical Methods in Natural Language Processing",
    year = "2024",
    pages = "17889--17904",
}

@inproceedings{4,
    title = "{ERD}: A Framework for Improving {LLM} Reasoning for Cognitive Distortion Classification",
    author = "Lim, Sehee  and
      Kim, Yejin  and
      Choi, Chi-Hyun  and
      Sohn, Jy-yong  and
      Kim, Byung-Hoon",
    booktitle = "Procs. of the 6th Clinical Natural Language Processing Workshop",
    year = "2024",
    pages = "292--300"
}

@INPROCEEDINGS{5,
  author={Huang, Ronghong},
  booktitle={Procs. of the IEEE 4th International Conference on Information Technology, Big Data and Artificial Intelligence}, 
  title={Improving Spam Detection with a Multi-Agent Debate Framework}, 
  year={2024},
  volume={4},
  number={},
}

@inproceedings{6,
author = {Smit, Andries and Grinsztajn, Nathan and Duckworth, Paul and Barrett, Thomas D. and Pretorius, Arnu},
title = {Should we be going MAD? a look at multi-agent debate strategies for LLMs},
year = {2024},
booktitle = {Procs. of the 41st International Conference on Machine Learning},
articleno = {1866},
numpages = {23},
location = {Vienna, Austria},
series = {ICML'24}
}

@INPROCEEDINGS{7,
  author={Bai, Yilin},
  booktitle={Procs. of the 10th International Conference on Big Data and Information Analytics}, 
  title={Confidence{C}al: Enhancing {LLM}s Reliability through Confidence Calibration in Multi-Agent Debate}, 
  year={2024},
  volume={},
  number={},
}

@article{8,
title = {Learning to break: Knowledge-enhanced reasoning in multi-agent debate system},
journal = {Neurocomputing},
volume = {618},
year = {2025},
author = {Haotian Wang and Xiyuan Du and Weijiang Yu and others}
}

@InProceedings{9,
author="Jeptoo, Korir Nancy
and Sun, Chengjie",
title="Enhancing Fake News Detection with Large Language Models Through Multi-agent Debates",
booktitle="Natural Language Processing and Chinese Computing",
year="2025"
}

@inproceedings{10,
    title = "Debate as Optimization: Adaptive Conformal Prediction and Diverse Retrieval for Event Extraction",
    author = "Wang, Sijia  and
      Huang, Lifu",
    booktitle = "Findings of the Association for Computational Linguistics: EMNLP 2024",
    year = "2024",
    pages = "16422--16435"
}

@inproceedings{11,
    title = "{C}o{E}vol: Constructing Better Responses for Instruction Finetuning through Multi-Agent Cooperation",
    author = "Li, Renhao  and
      Tan, Minghuan  and
      Wong, Derek F.  and
      Yang, Min",
    booktitle = "Procs. of the 2024 Conference on Empirical Methods in Natural Language Processing",
    year = "2024",
    pages = "4703--4721"
}

@inproceedings{12,
title={Chat{E}val: Towards Better {LLM}-based Evaluators through Multi-Agent Debate},
author={Chi-Min Chan and Weize Chen and Yusheng Su and others},
booktitle={Procs. of the 12th International Conference on Learning Representations},
year={2024}
}

@inproceedings{14,
title={Confidence Calibration and Rationalization for {LLM}s via Multi-Agent Deliberation},
author={Ruixin Yang and Dheeraj Rajagopal and Shirley Anugrah Hayati and Bin Hu and Dongyeop Kang},
booktitle={ICLR 2024 Workshop on Reliable and Responsible Foundation Models},
year={2024}
}

@misc{15,
      title={Enhancing Diagnostic Accuracy through Multi-Agent Conversations: Using Large Language Models to Mitigate Cognitive Bias}, 
      author={Yu He Ke and Rui Yang and Sui An Lie and others},
  archivePrefix = {arXiv},
  eprint = {2401.14589},
  journal = {arXiv},
      year={2024},
}

@inproceedings{16,
    title = "Examining Inter-Consistency of Large Language Models Collaboration: An In-depth Analysis via Debate",
    author = "Xiong, Kai  and
      Ding, Xiao  and
      Cao, Yixin  and
      Liu, Ting  and
      Qin, Bing",
    booktitle = "Findings of the Association for Computational Linguistics: EMNLP 2023",
    year = "2023",
    pages = "7572--7590"}

@inproceedings{17,
title={Let Models Speak Ciphers: Multiagent Debate through Embeddings},
author={Chau Pham and Boyi Liu and Yingxiang Yang and Zhengyu Chen and Tianyi Liu and Jianbo Yuan and Bryan A. Plummer and Zhaoran Wang and Hongxia Yang},
booktitle={The Twelfth International Conference on Learning Representations},
year={2024}
}

@inproceedings{18,
title={Improving Factuality and Reasoning in Language Models through Multiagent Debate},
author={Yilun Du and Shuang Li and Antonio Torralba and Joshua B. Tenenbaum and Igor Mordatch},
booktitle={Procs. of the 41st International Conference on Machine Learning},
year={2024}
}

@misc{19,
      title={{Multi-Agent Consensus Seeking via Large Language Models}}, 
      author={Huaben Chen and Wenkang Ji and Lufeng Xu and Shiyu Zhao},
      year={2025},
      eprint={2310.20151},
      archivePrefix={arXiv},
      primaryClass={cs.CL},
      url={https://arxiv.org/abs/2310.20151}, 
      note = {arXiv preprint arXiv:2310.20151}
}

@article{20,
title={{PRD}: Peer Rank and Discussion Improve Large Language Model based Evaluations},
author={Ruosen Li and Teerth Patel and Xinya Du},
journal={Transactions on Machine Learning Research},
issn={2835-8856},
year={2024},
note={}
}

@inproceedings{21,
    title = "{R}e{C}oncile: Round-Table Conference Improves Reasoning via Consensus among Diverse {LLM}s",
    author = "Chen, Justin  and
      Saha, Swarnadeep  and
      Bansal, Mohit",
    booktitle = "Procs. of the 62nd Annual Meeting of the Association for Computational Linguistics (Volume 1: Long Papers)",
    year = "2024",
}

@inproceedings{22,
title = "Rethinking the Bounds of LLM Reasoning: Are Multi-Agent Discussions the Key?",
author = "Qineng Wang and Zihao Wang and Ying Su and Hanghang Tong and Yangqiu Song",
year = "2024",
booktitle = "Procs. of the Annual Meeting of the Association for Computational Linguistics",
pages = "6106--6131"
}

@inproceedings{23,
  author={Xinyu Zhu and Junjie Wang and Lin Zhang and Yuxiang Zhang and Yongfeng Huang and Ruyi Gan and Jiaxing Zhang and Yujiu Yang},
  title={Solving Math Word Problems via Cooperative Reasoning induced Language Models},
  year={2023},
  pages={4471-4485},
  booktitle={ACL (1)}
}

@misc{24,
      title={Towards Reasoning in Large Language Models via Multi-Agent Peer Review Collaboration}, 
      author={Zhenran Xu and Senbao Shi and Baotian Hu and Jindi Yu and Dongfang Li and Min Zhang and Yuxiang Wu},
      year={2023},
      eprint={2311.08152},
      archivePrefix={arXiv},
      primaryClass={cs.CL},
      url={https://arxiv.org/abs/2311.08152}, 
      note = {arXiv preprint arXiv:2311.08152}
}

@inproceedings{25,
    title = "Unleashing the Emergent Cognitive Synergy in Large Language Models: A Task-Solving Agent through Multi-Persona Self-Collaboration",
    author = "Wang, Zhenhailong  and
      Mao, Shaoguang  and
      Wu, Wenshan  and
      Ge, Tao  and
      Wei, Furu  and
      Ji, Heng",
    booktitle = "Procs. of the 2024 Conference of the North American Chapter of the Association for Computational Linguistics: Human Language Technologies (Volume 1: Long Papers)",
    year = "2024",
    pages = "257--279"
}

@inproceedings{26,
  author = {Zijun Liu and Yanzhe Zhang and Peng Li and Yang Liu and Diyi Yang},
  title = {A dynamic {LLM}-powered agent network for task-oriented agent collaboration},
  year = {2024},
  booktitle = {Procs. of the 1st Conference on Language Modeling}
}

@article{27,
  author = {Guoxi Zhang and Jiawei Chen and Tianzhuo Yang and Jiaming Ji and Yaodong Yang and Juntao Dai},
  title = {A Game-Theoretic Negotiation Framework for Cross-Cultural Consensus in LLMs},
  year = {2025},
  archivePrefix = {arXiv},
  eprint = {2506.13245},
  journal = {arXiv},
  note = {arXiv preprint arXiv:2506.13245}
}

@article{28,
  author = {Andrew Estornell and Jean-François Ton and Yuanshun Yao and Yang Liu},
  title = {{ACC}-collab: An actor-critic approach to multi-agent {LLM} collaboration},
  year = {2024},
  archivePrefix = {arXiv},
  eprint = {2411.00053},
  journal = {arXiv},
  note = {arXiv preprint arXiv:2411.00053}
}

@article{29,
  author = {Linxin Song and Jiale Liu and Jieyu Zhang and Shaokun Zhang and Ao Luo and Shijian Wang and Qingyun Wu and Chi Wang},
  title = {Adaptive in-conversation team building for language model agents},
  year = {2024},
  archivePrefix = {arXiv},
  eprint = {2405.19425},
  journal = {arXiv},
  note = {arXiv preprint arXiv:2405.19425}
}

@inproceedings{30,
    title = "{A}gent{D}ropout: Dynamic Agent Elimination for Token-Efficient and High-Performance {LLM}-Based Multi-Agent Collaboration",
    author = "Wang, Zhexuan  and
      Wang, Yutong  and
      Liu, Xuebo  and
      others",
    booktitle = "Procs. of the 63rd Annual Meeting of the Association for Computational Linguistics (Volume 1: Long Papers)",
    year = "2025",
    publisher = "ACL",
    url = "https://aclanthology.org/2025.acl-long.1170/",
    doi = "10.18653/v1/2025.acl-long.1170",
    pages = "24013--24035",
    ISBN = "979-8-89176-251-0"
}

@inproceedings{31,
    title = "An Empirical Study of Group Conformity in Multi-Agent Systems",
    author = "Choi, Min  and
      Kim, Keonwoo  and
      Chae, Sungwon  and
      Baek, Sangyeop",
    booktitle = "Findings of the Association for Computational Linguistics: ACL 2025",
    year = "2025",
    publisher = "ACL",
    url = "https://aclanthology.org/2025.findings-acl.265/",
    doi = "10.18653/v1/2025.findings-acl.265",
    ISBN = "979-8-89176-256-5"
}

@inproceedings{32,
  author = {Andries Petrus Smit and Paul Duckworth and Nathan Grinsztajn and Kale-ab Tessera and Thomas D Barrett and Arnu Pretorius},
  title = {Are we going mad? benchmarking multi-agent debate between language models for medical q\&a},
  year = {2023},
  booktitle = {Deep Generative Models for Health Workshop NeurIPS 2023}
}

@inproceedings{33,
  author = {So Watanabe and Chee Siang Leow and Junichi Hoshino and others},
  title = {Assessment and Improvement of Customer Service Speech with Multiple Large Language Models},
  year = {2024},
  booktitle = {Procs. of the 2024 Asia Pacific Signal and Information Processing Association Annual Summit and Conference},
  publisher = {IEEE}
}

@inproceedings{35,
    title = "Auto-Arena: Automating {LLM} Evaluations with Agent Peer Battles and Committee Discussions",
    author = "Zhao, Ruochen  and
      Zhang, Wenxuan  and
      Chia, Yew Ken  and
      Xu, Weiwen  and
      Zhao, Deli  and
      Bing, Lidong",
    booktitle = "Procs. of the 63rd Annual Meeting of the Association for Computational Linguistics (Volume 1: Long Papers)",
    year = "2025",
    publisher = "ACL",
    url = "https://aclanthology.org/2025.acl-long.223/",
    doi = "10.18653/v1/2025.acl-long.223",
    pages = "4440--4463",
    ISBN = "979-8-89176-251-0"
}

@inproceedings{36,
    title = "{BELLE}: A Bi-Level Multi-Agent Reasoning Framework for Multi-Hop Question Answering",
    author = "Zhang, Taolin  and
      Li, Dongyang  and
      Chen, Qizhou  and
      Wang, Chengyu  and
      He, Xiaofeng",
    booktitle = "Procs. of the 63rd Annual Meeting of the Association for Computational Linguistics (Volume 1: Long Papers)",
    year = "2025",
    publisher = "ACL",
    url = "https://aclanthology.org/2025.acl-long.211/",
    doi = "10.18653/v1/2025.acl-long.211",
    pages = "4184--4202",
    ISBN = "979-8-89176-251-0"
}

@inproceedings{37,
    title = "Beyond Frameworks: Unpacking Collaboration Strategies in Multi-Agent Systems",
    author = "Wang, Haochun  and
      Zhao, Sendong  and
      Wang, Jingbo  and
      Qiang, Zewen  and
      Qin, Bing  and
      Liu, Ting",
    booktitle = "Procs. of the 63rd Annual Meeting of the Association for Computational Linguistics (Volume 1: Long Papers)",
    year = "2025",
    publisher = "ACL",
    url = "https://aclanthology.org/2025.acl-long.1037/",
    doi = "10.18653/v1/2025.acl-long.1037",
    pages = "21361--21375",
    ISBN = "979-8-89176-251-0"
}

@INPROCEEDINGS{38,
  author={Qin, Zining and Wang, Chenhao and Guo, Jianxiong and Qin, Huiling and Jia, Weijia},
  booktitle={IEEE International Conference on Multimedia and Expo}, 
  title={Brainstorming Brings Power to Large Language Models of Knowledge Reasoning}, 
  year={2025},
  volume={},
  number={},
  pages={1-6},
  doi={10.1109/ICME59968.2025.11209475}}

@article{39,
  author = {Mingqing Zhang and Haisong Gong and Qiang Liu and Shu Wu and Liang Wang},
  title = {Breaking event rumor detection via stance-separated multi-agent debate},
  year = {2024},
  archivePrefix = {arXiv},
  eprint = {2412.04859},
  journal = {arXiv},
  note = {arXiv preprint arXiv:2412.04859}
}

@inproceedings{40,
  author = {Yexiang Liu and Jie Cao and Zekun Li and Ran He and Tieniu Tan},
  title = {Breaking mental set to improve reasoning through diverse multi-agent debate},
  year = {2025},
  booktitle = {Procs. of the 13th International Conference on Learning Representations}
}

@INPROCEEDINGS{41,
  author={Zhao, Yiheng and Yan, Jun},
  booktitle={2025 IEEE Symposium on Trustworthy, Explainable and Responsible Computational Intelligence (CITREx)}, 
  title={Can LLMs Identify Event Causality More Accurately through Debate? A Systematic Assessment of LLMs’ Factuality and Reasoning}, 
  year={2025},
  volume={},
  number={},
  pages={1-8},
  doi={10.1109/CITREx64975.2025.10974935}}

@inproceedings{43,
author = {Park, Jeongeon and Min, Bryan and Son, Kihoon and Song, Jean Y and Ma, Xiaojuan and Kim, Juho},
title = {ChoiceMates: Supporting Unfamiliar Online Decision-Making with Multi-Agent Conversational Interactions},
year = {2026},
isbn = {9798400719844},
publisher = {ACM},
url = {https://doi.org/10.1145/3742413.3789107},
doi = {10.1145/3742413.3789107},
booktitle = {Procs. of the 31st International Conference on Intelligent User Interfaces},
pages = {1526–1550},
numpages = {25},
}

@inproceedings{44,
  author = {Zixiang Wang and Yinghao Zhu and Huiya Zhao and Xiaochen Zheng and Dehao Sui and Tianlong Wang and Wen Tang and Yasha Wang and Ewen Harrison and Chengwei Pan},
  title = {Colacare: Enhancing electronic health record modeling through large language model-driven multi-agent collaboration},
  year = {2025},
  booktitle = {Procs. of the ACM on Web Conference},
  pages = {2250-2261}
}

@inproceedings{45,
    title = "{CONSENSAGENT}: Towards Efficient and Effective Consensus in Multi-Agent {LLM} Interactions Through Sycophancy Mitigation",
    author = "Pitre, Priya  and
      Ramakrishnan, Naren  and
      Wang, Xuan",
    booktitle = "Findings of the Association for Computational Linguistics: ACL 2025",
    year = "2025",
    publisher = "ACL",
    doi = "10.18653/v1/2025.findings-acl.1141",
    pages = "22112--22133"
}

@inproceedings{46,
  author = {Kaiqi Yang and Yucheng Chu and Taylor Darwin and others},
  title = {Content knowledge identification with multi-agent large language models (LLMs)},
  year = {2024},
  booktitle = {International Conference on Artificial Intelligence in Education},
  pages = {284-292},
  publisher = {Springer}
}

@article{48,
  author = {Qiushi Sun and Zhangyue Yin and Xiang Li and Zhiyong Wu and Xipeng Qiu and Lingpeng Kong},
  title = {Corex: Pushing the boundaries of complex reasoning through multi-model collaboration},
  year = {2023},
  archivePrefix = {arXiv},
  eprint = {2310.00280},
  journal = {arXiv},
  note = {arXiv preprint arXiv:2310.00280}
}

@inproceedings{49,
    title = "{C}ortex{D}ebate: Debating Sparsely and Equally for Multi-Agent Debate",
    author = "Sun, Yiliu  and
      Zhao, Zicheng  and
      Wan, Sheng  and
      Gong, Chen",
    booktitle = "Findings of the Association for Computational Linguistics: ACL 2025",
    year = "2025",
    url = "https://aclanthology.org/2025.findings-acl.495/",
    doi = "10.18653/v1/2025.findings-acl.495",
    pages = "9503--9523",
    ISBN = "979-8-89176-256-5"
}

@article{50,
  author = {Weiqiang Jin and Dafu Su and Tao Tao and others},
  title = {Courtroom-{FND}: a multi-role fake news detection method based on argument switching-based courtroom debate},
  year = {2025},
  journal = {Journal of King Saud University Computer and Information Sciences},
  number = {3},
  pages = {33},
  publisher = {Springer},
  volume = {37.0}
}

@inproceedings{51,
    title = "Can an Individual Manipulate the Collective Decisions of Multi-Agents?",
    author = "Liu, Fengyuan  and
      Zhao, Rui  and
      Chen, Shuo  and
      others",
    booktitle = "Procs. of the 2025 Conference on Empirical Methods in Natural Language Processing",
    year = "2025",
    publisher = "ACL",
    url = "https://aclanthology.org/2025.emnlp-main.611/",
    doi = "10.18653/v1/2025.emnlp-main.611",
    pages = "12158--12182",
    ISBN = "979-8-89176-332-6"
}

@article{52,
  author = {Guibin Zhang and Yanwei Yue and Zhixun Li and Sukwon Yun and Guancheng Wan and Kun Wang and Dawei Cheng and Jeffrey Xu Yu and Tianlong Chen},
  title = {Cut the crap: An economical communication pipeline for llm-based multi-agent systems},
  year = {2024},
  archivePrefix = {arXiv},
  eprint = {2410.02506},
  journal = {arXiv},
  note = {arXiv preprint arXiv:2410.02506}
}

@inproceedings{54,
    title = "Debate, Reflect, and Distill: Multi-Agent Feedback with Tree-Structured Preference Optimization for Efficient Language Model Enhancement",
    author = "Zhou, Xiaofeng  and
      Huang, Heyan  and
      Liao, Lizi",
    booktitle = "Findings of the ACL 2025",
    year = "2025",
    publisher = "ACL",
    url = "https://aclanthology.org/2025.findings-acl.475/",
    doi = "10.18653/v1/2025.findings-acl.475",
    pages = "9122--9137",
    ISBN = "979-8-89176-256-5"
}

@inproceedings{55,
    title = "{DEBATE}, {TRAIN}, {EVOLVE}: {S}elf{-}{E}volution of Language Model Reasoning",
    author = "Srivastava, Gaurav  and
      Bi, Zhenyu  and
      Lu, Meng  and
      Wang, Xuan",
    booktitle = "Procs. of the 2025 Conference on Empirical Methods in Natural Language Processing",
    year = "2025",
    publisher = "ACL",
    url = "https://aclanthology.org/2025.emnlp-main.1666/",
    doi = "10.18653/v1/2025.emnlp-main.1666",
    pages = "32764--32810",
    ISBN = "979-8-89176-332-6"
}

@inproceedings{56,
    title = "{DEBATE}: Devil{'}s Advocate-Based Assessment and Text Evaluation",
    author = "Kim, Alex  and
      Kim, Keonwoo  and
      Yoon, Sangwon",
    booktitle = "Findings of the Association for Computational Linguistics: ACL 2024",
    year = "2024",
    publisher = "ACL",
    url = "https://aclanthology.org/2024.findings-acl.112/",
    doi = "10.18653/v1/2024.findings-acl.112"
}

@inproceedings{57,
  author = {Ngoc Tuong Vy Nguyen and Felix D Childress and Yunting Yin},
  title = {Debate-Driven Multi-Agent LLMs for Phishing Email Detection},
  year = {2025},
  booktitle = {Procs. of the 13th International Symposium on Digital Forensics and Security},
  publisher = {IEEE}
}

@inproceedings{58,
    title = "Debate-to-Detect: Reformulating Misinformation Detection as a Real-World Debate with Large Language Models",
    author = "Han, Chen  and
      Zheng, Wenzhen  and
      Tang, Xijin",
    booktitle = "Procs. of the 2025 Conference on Empirical Methods in Natural Language Processing",
    year = "2025",
    publisher = "ACL",
    url = "https://aclanthology.org/2025.emnlp-main.764/",
    doi = "10.18653/v1/2025.emnlp-main.764",
    ISBN = "979-8-89176-332-6"
}

@inproceedings{59,
    title = "Debate-to-Write: A Persona-Driven Multi-Agent Framework for Diverse Argument Generation",
    author = "Hu, Zhe  and
      Chan, Hou Pong  and
      Li, Jing  and
      Yin, Yu",
    booktitle = "Procs. of the 31st International Conference on Computational Linguistics",
    year = "2025",
    publisher = "ACL",
    url = "https://aclanthology.org/2025.coling-main.314/"
}

@inproceedings{60,
    title = "{D}eb{U}nc: Improving Large Language Model Agent Communication With Uncertainty Metrics",
    author = "Yoffe, Luke  and
      Amayuelas, Alfonso  and
      Wang, William Yang",
    booktitle = "Findings of the Association for Computational Linguistics: EMNLP 2025",
    year = "2025",
   publisher = "ACL",
    url = "https://aclanthology.org/2025.findings-emnlp.1265/",
    doi = "10.18653/v1/2025.findings-emnlp.1265",
    pages = "23299--23315",
    ISBN = "979-8-89176-335-7"
}

@article{61,
  author = {G Bharathi Mohan and M Gayathri and R Prasanna Kumar},
  title = {Detoxifying language model outputs: combining multi-agent debates and reinforcement learning for improved summarization},
  year = {2025},
  journal = {Language Resources and Evaluation},
  publisher = {Springer}
}

@inproceedings{62,
    title = "Do Androids Question Electric Sheep? A Multi-Agent Cognitive Simulation of Philosophical Reflection on Hybrid Table Reasoning",
    author = "Ma, Yiran Rex",
    booktitle = "Procs. of the 63rd Annual Meeting of the Association for Computational Linguistics (Volume 4: Student Research Workshop)",
    year = "2025",
    publisher = "ACL",
    doi = "10.18653/v1/2025.acl-srw.9",
    pages = "143--164"
}

@article{63,
  author = {Yue Cui and Liuyi Yao and Zitao Li and Yaliang Li and Bolin Ding and Xiaofang Zhou},
  title = {Efficient Leave-one-out Approximation in {LLM} Multi-agent Debate Based on Introspection},
  year = {2025},
  archivePrefix = {arXiv},
  eprint = {2505.22192},
  journal = {arXiv},
  note = {arXiv preprint arXiv:2505.22192}
}

@article{64,
  author = {Xi Chen and Huahui Yi and Mingke You and others},
  title = {Enhancing diagnostic capability with multi-agents conversational large language models},
  year = {2025},
  journal = {NPJ digital medicine},
  number = {1},
  publisher = {Nature Publishing Group UK London},
  volume = {8.0}
}

@inproceedings{65,
  author = {Yiyue Qian and Shinan Zhang and Yun Zhou and Haibo Ding and Diego Socolinsky and Yi Zhang},
  title = {Enhancing {LLM}-as-a-judge via multi-agent collaboration},
  year = {2025},
  booktitle = {Procs. of the AAAI 2025 Workshop on Advancing LLM-Based Multi-Agent Collaboration}
}

@inproceedings{66,
  author = {Zhihua Duan and Jialin Wang},
  title = {Enhancing multi-agent consensus through third-party {LLM} integration: Analyzing uncertainty and mitigating hallucinations in large language models},
  year = {2025},
  booktitle = {Procs. of the 8th International Conference on Advanced Algorithms and Control Engineering},
  publisher = {IEEE}
}

@article{67,
  author = {Xin-Cheng Wen and Jiaxin Ye and Cuiyun Gao and Lianwei Wu and Qing Liao},
  title = {EvalSVA: Multi-Agent Evaluators for Next-Gen Software Vulnerability Assessment},
  year = {2024},
  archivePrefix = {arXiv},
  eprint = {2501.14737},
  journal = {arXiv},
  note = {arXiv preprint arXiv:2501.14737}
}

@inproceedings{68,
  author = {Yidong He and Yongbin Liu and Chunping Ouyang and others},
  title = {Evaluating Human-Large Language Model Alignment in Group Process},
  year = {2024},
  booktitle = {CCF International Conference on Natural Language Processing and Chinese Computing},
  publisher = {Springer}
}

@inproceedings{69,
    title = "Evaluating the Performance of Large Language Models via Debates",
    author = "Moniri, Behrad  and
      Hassani, Hamed  and
      Dobriban, Edgar",
    booktitle = "Findings of the Association for Computational Linguistics: NAACL 2025",
    year = "2025",
    publisher = "ACL",
    url = "https://aclanthology.org/2025.findings-naacl.109/",
    doi = "10.18653/v1/2025.findings-naacl.109",
    pages = "2040--2075",
    ISBN = "979-8-89176-195-7"
}

@inbook{70,
title = {EVINCE: Optimizing Adversarial LLM Dialogues via Conditional Statistics and Information Theory},
      author={Edward Y. Chang},
      year = {2025},
isbn = {9798400731976},
publisher = {ACM},
url = {https://doi.org/10.1145/3749421.3749431},
booktitle = {Multi-LLM Agent Collaborative Intelligence: The Path to Artificial General Intelligence}
}

@inproceedings{72,
    title = "Exchange-of-Thought: Enhancing Large Language Model Capabilities through Cross-Model Communication",
    author = "Yin, Zhangyue  and
      Sun, Qiushi  and
      Chang, Cheng  and
      others",
    booktitle = "Procs. of the 2023 Conference on Empirical Methods in Natural Language Processing",
    year = "2023",
    publisher = "ACL",
    url = "https://aclanthology.org/2023.emnlp-main.936/",
    doi = "10.18653/v1/2023.emnlp-main.936",
    pages = "15135--15153"
}

@inproceedings{74,
    title = "Exploring Collaboration Mechanisms for {LLM} Agents: A Social Psychology View",
    author = "Zhang, Jintian  and
      Xu, Xin  and
      Zhang, Ningyu  and
      others",
    booktitle = "Procs. of the 62nd Annual Meeting of the Association for Computational Linguistics (Volume 1: Long Papers)",
    year = "2024",
    publisher = "ACL",
    url = "https://aclanthology.org/2024.acl-long.782/",
    doi = "10.18653/v1/2024.acl-long.782",
    pages = "14544--14607"
}

@article{75,
  author = {Junxia Ma and Changjiang Wang and Lu Rong and Bo Wang and Yaoli Xu},
  title = {Exploring multi-agent debate for zero-shot stance detection: A novel approach},
  year = {2025},
  journal = {Applied Sciences},
  number = {9},
  pages = {4612},
  publisher = {MDPI},
  volume = {15.0}
}

@inproceedings{76,
    title = "Faithful, Unfaithful or Ambiguous? Multi-Agent Debate with Initial Stance for Summary Evaluation",
    author = "Koupaee, Mahnaz  and
      Vincent, Jake W.  and
      Mansour, Saab  and
      others",
    booktitle = "Procs. of the 2025 Conference of the Nations of the Americas Chapter of the Association for Computational Linguistics: Human Language Technologies (Volume 1: Long Papers)",
    year = "2025",
    publisher = "ACL",
    url = "https://aclanthology.org/2025.naacl-long.609/",
    doi = "10.18653/v1/2025.naacl-long.609",
    ISBN = "979-8-89176-189-6"
}

@article{78,
  author = {Masahiro Sato},
  title = {{GAI}: Generative Agents for Innovation},
  year = {2024},
  archivePrefix = {arXiv},
  eprint = {2412.18899},
  journal = {arXiv},
  note = {arXiv preprint arXiv:2412.18899}
}

@article{80,
  author = {Rui Zou and Mengqi Wei and Jintian Feng and Qian Wan and Jianwen Sun and Sannyuya Liu},
  title = {Gradual vigilance and interval communication: Enhancing value alignment in multi-agent debates},
  year = {2024},
  archivePrefix = {arXiv},
  eprint = {2412.13471},
  journal = {arXiv},
  note = {arXiv preprint arXiv:2412.13471}
}

@article{81,
  author = {Tongxuan Liu and Xingyu Wang and Weizhe Huang and others},
  title = {GroupDebate: Enhancing the efficiency of multi-agent debate using group discussion},
  year = {2024},
  archivePrefix = {arXiv},
  eprint = {2409.14051},
  journal = {arXiv},
  note = {arXiv preprint arXiv:2409.14051}
}

@InProceedings{82,
author="Li, Yu
and Huang, Yi
and Qi, Guilin
and others",
title="Harnessing Diverse Perspectives: A Multi-agent Framework for Enhanced Error Detection in Knowledge Graphs",
booktitle="Database Systems for Advanced Applications",
year="2026",
publisher="Springer",
pages="446--458",
isbn="978-981-95-4158-4"
}

@article{85,
  author = {Baiting Chen and Tong Zhu and Jiale Han and others},
  title = {Incentivizing Truthful Language Models via Peer Elicitation Games},
  year = {2025},
  archivePrefix = {arXiv},
  eprint = {2505.13636},
  journal = {arXiv},
  note = {arXiv preprint arXiv:2505.13636}
}

@inproceedings{87,
    title = "Into the Unknown Unknowns: Engaged Human Learning through Participation in Language Model Agent Conversations",
    author = "Jiang, Yucheng  and
      Shao, Yijia  and
      Ma, Dekun  and
      Semnani, Sina  and
      Lam, Monica",
    booktitle = "Procs. of the 2024 Conference on Empirical Methods in Natural Language Processing",
    year = "2024",
    publisher = "ACL",
    url = "https://aclanthology.org/2024.emnlp-main.554/",
    doi = "10.18653/v1/2024.emnlp-main.554"
}

@article{88,
  author = {Jina Chun and Qihong Chen and Jiawei Li and Iftekhar Ahmed},
  title = {Is multi-agent debate ({MAD}) the silver bullet? An empirical analysis of {MAD} in code summarization and translation},
  year = {2025},
  archivePrefix = {arXiv},
  eprint = {2503.12029},
  journal = {arXiv},
  note = {arXiv preprint arXiv:2503.12029}
}

@inproceedings{90,
    title = "Judging with Many Minds: Do More Perspectives Mean Less Prejudice? On Bias Amplification and Resistance in Multi-Agent Based {LLM}-as-Judge",
    author = "Ma, Chiyu  and
      Zhang, Enpei  and
      Zhao, Yilun  and
      others",
    booktitle = "Findings of the Association for Computational Linguistics",
    year = "2025",
    publisher = "ACL",
    url = "https://aclanthology.org/2025.findings-emnlp.941/",
    doi = "10.18653/v1/2025.findings-emnlp.941",
    pages = "17356--17392",
    ISBN = "979-8-89176-335-7"
}

@article{91,
  author = {Yuran Li and Jama Hussein Mohamud and Chongren Sun and Di Wu and Benoit Boulet},
  title = {Leveraging {LLM}s as meta-judges: A multi-agent framework for evaluating {LLM} judgments},
  year = {2025},
  archivePrefix = {arXiv},
  eprint = {2504.17087},
  journal = {arXiv},
  note = {arXiv preprint arXiv:2504.17087}
}

@article{92,
  author = {Li-Chun Lu and Shou-Jen Chen and Tsung-Min Pai and others},
  title = {{LLM} discussion: Enhancing the creativity of large language models via discussion framework and role-play},
  year = {2024},
  archivePrefix = {arXiv},
  eprint = {2405.06373},
  journal = {arXiv},
  note = {arXiv preprint arXiv:2405.06373}
}

@inproceedings{93,
author = {Chong, Zan-Kai and Ohsaki, Hiroyuki and Ng, Bryan},
title = {{LLM}-Net: Democratizing {LLM}s-as-a-Service through Blockchain-based Expert Networks},
year = {2025},
isbn = {9798400710124},
publisher = {ACM},
url = {https://doi.org/10.1145/3731806.3731827},
doi = {10.1145/3731806.3731827},
booktitle = {Procs. of the 14th International Conference on Software and Computer Applications},
numpages = {8},
}

@ARTICLE{95,
  author={Wei, Zhiyuan and Sun, Jing and Sun, Yuqiang and others},
  journal={IEEE Transactions on Software Engineering}, 
  title={Advanced Smart Contract Vulnerability Detection via LLM-Powered Multi-Agent Systems}, 
  year={2025},
  volume={51},
  number={10},
  pages={2830-2846},
  doi={10.1109/TSE.2025.3597319}}

@inproceedings{96,
author = {Chen, Justin Chih-Yao and Saha, Swarnadeep and Stengel-Eskin, Elias and Bansal, Mohit},
title = {{MAGDI}: structured distillation of multi-agent interaction graphs improves reasoning in smaller language models},
year = {2024},
booktitle = {Procs. of the 41st International Conference on Machine Learning},
articleno = {280},
numpages = {16}
}

@inproceedings{98,
    title = "{MAMM}-Refine: A Recipe for Improving Faithfulness in Generation with Multi-Agent Collaboration",
    author = "Wan, David  and
      Chen, Justin  and
      Stengel-Eskin, Elias  and
      Bansal, Mohit",
    booktitle = "Procs. of the 2025 Conference of the Nations of the Americas Chapter of the Association for Computational Linguistics: Human Language Technologies (Volume 1: Long Papers)",
    year = "2025",
    publisher = "ACL",
    url = "https://aclanthology.org/2025.naacl-long.498/",
    doi = "10.18653/v1/2025.naacl-long.498",
    pages = "9882--9901",
    ISBN = "979-8-89176-189-6"
}

@inproceedings{99,
    title = "{MAP}o{RL}: Multi-Agent Post-Co-Training for Collaborative Large Language Models with Reinforcement Learning",
    author = "Park, Chanwoo  and
      Han, Seungju  and
      Guo, Xingzhi  and
      others",
    booktitle = "Procs. of the 63rd Annual Meeting of the Association for Computational Linguistics (Volume 1: Long Papers)",
    year = "2025",
    publisher = "ACL",
    url = "https://aclanthology.org/2025.acl-long.1459/",
    doi = "10.18653/v1/2025.acl-long.1459",
    ISBN = "979-8-89176-251-0"
}

@InProceedings{101,
author="Li, Yu
and Zhang, Shenyu
and Wu, Rui
and others",
title="MATEval: A Multi-agent Discussion Framework for Advancing Open-Ended Text Evaluation",
booktitle="Database Systems for Advanced Applications",
year="2024",
publisher="Springer",
pages="415--426",
isbn="978-981-97-5575-2"
}

@article{103,
  author = {Yeonji Lee and Sangjun Park and Kyunghyun Cho and JinYeong Bak},
  title = {MentalAgora: A Gateway to Advanced Personalized Care in Mental Health through Multi-Agent Debating and Attribute Control},
  year = {2024},
  archivePrefix = {arXiv},
  eprint = {2407.02736},
  journal = {arXiv},
  note = {arXiv preprint arXiv:2407.02736}
}

@article{105,
  author = {Yi Yang and Yitong Ma and Hao Feng and Yiming Cheng and Zhu Han},
  title = {Minimizing hallucinations and communication costs: Adversarial debate and voting mechanisms in llm-based multi-agents},
  year = {2025},
  journal = {Applied Sciences},
  number = {7},
  pages = {3676},
  publisher = {MDPI},
  volume = {15.0}
}

@article{106,
  author = {Li Zhang and Kevin D Ashley},
  title = {Mitigating Manipulation and Enhancing Persuasion: A Reflective Multi-Agent Approach for Legal Argument Generation},
  year = {2025},
  archivePrefix = {arXiv},
  eprint = {2506.02992},
  journal = {arXiv},
  note = {arXiv preprint arXiv:2506.02992}
}

@inproceedings{107,
    title = "{M}-{MAD}: Multidimensional Multi-Agent Debate for Advanced Machine Translation Evaluation",
    author = "Feng, Zhaopeng  and
      Su, Jiayuan  and
      Zheng, Jiamei  and
      others",
    booktitle = "Procs. of the 63rd Annual Meeting of the Association for Computational Linguistics (Volume 1: Long Papers)",
    year = "2025",
    publisher = "ACL",
    url = "https://aclanthology.org/2025.acl-long.351/",
    doi = "10.18653/v1/2025.acl-long.351",
    pages = "7084--7107",
    ISBN = "979-8-89176-251-0"
}

@inproceedings{108,
  author = {Yong Guan and Hao Peng and Lei Hou and Juanzi Li},
  title = {Mmd-ere: multi-agent multi-sided debate for event relation extraction},
  year = {2025},
  booktitle = {Procs. of the 31st International Conference on Computational Linguistics},
}

@article{109,
  author = {Farhad Moghimifar and Yuan-Fang Li and Robert Thomson and Gholamreza Haffari},
  title = {Modelling political coalition negotiations using llm-based agents},
  year = {2024},
  archivePrefix = {arXiv},
  eprint = {2402.11712},
  journal = {arXiv},
  note = {arXiv preprint arXiv:2402.11712}
}

@article{111,
  author = {Hao Duong Le and Xin Xia and Zhang Chen},
  title = {Multi-agent causal discovery using large language models},
  year = {2024},
  archivePrefix = {arXiv},
  eprint = {2407.15073},
  journal = {arXiv},
  note = {arXiv preprint arXiv:2407.15073}
}

@inproceedings{114,
author = {Estornell, Andrew and Liu, Yang},
title = {Multi-{LLM} debate: framework, principals, and interventions},
year = {2024},
isbn = {9798331314385},
publisher = {Curran Associates Inc.},
booktitle = {Procs. of the 38th International Conference on Neural Information Processing Systems},
articleno = {911},
numpages = {27},
}

@article{115,
  author = {Bhrij Patel and Vishnu Sashank Dorbala and Amrit Singh Bedi and Dinesh Manocha},
  title = {Multi-{LLM} {QA} with Embodied Exploration},
  year = {2024},
  archivePrefix = {arXiv},
  eprint = {2406.10918},
  journal = {arXiv},
  note = {arXiv preprint arXiv:2406.10918}
}

@inproceedings{116,
    title = "Multiple {LLM} Agents Debate for Equitable Cultural Alignment",
    author = "Ki, Dayeon  and
      Rudinger, Rachel  and
      Zhou, Tianyi  and
      Carpuat, Marine",
    booktitle = "Procs. of the 63rd Annual Meeting of the Association for Computational Linguistics (Volume 1: Long Papers)",
    year = "2025",
    publisher = "ACL",
    url = "https://aclanthology.org/2025.acl-long.1210/",
    doi = "10.18653/v1/2025.acl-long.1210",
    ISBN = "979-8-89176-251-0"
}

@article{117,
  author = {Zachary Kenton and Noah Siegel and János Kramár and others},
  title = {On scalable oversight with weak {LLM}s judging strong {LLM}s},
  year = {2024},
  journal = {Advances in Neural Information Processing Systems},
  volume = {37.0}
}

@article{118,
  author = {Qineng Wang and Zihao Wang and Ying Su and Yangqiu Song},
  title = {On the Discussion of Large Language Models: Symmetry of Agents and Interplay with Prompts},
  year = {2023},
  archivePrefix = {arXiv},
  eprint = {2311.07076},
  journal = {arXiv},
  note = {arXiv preprint arXiv:2311.07076}
}

@article{120,
  author = {Zhigeng Pan and Xianliang Xia and Fuchang Liu and Minglang Zheng},
  title = {PCcGE: Personalized Chinese Couplet Generation and Evaluation Framework Based on Large Language Models},
  year = {2025},
  journal = {Applied Sciences},
  number = {9},
  pages = {4996},
  publisher = {MDPI},
  volume = {15.0}
}

@article{121,
  author = {Dezheng Bao and Yueci Yang and Xin Chen and others},
  title = {{PD} $^ 3$: A Project Duplication Detection Framework via Adapted Multi-Agent Debate},
  year = {2025},
  archivePrefix = {arXiv},
  eprint = {2505.17492},
  journal = {arXiv},
  note = {arXiv preprint arXiv:2505.17492}
}

@article{122,
  author = {Wenhao Li and Selvakumar Manickam and Yung-wey Chong and Shankar Karuppayah},
  title = {PhishDebate: An {LLM}-Based Multi-Agent Framework for Phishing Website Detection},
  year = {2025},
  archivePrefix = {arXiv},
  eprint = {2506.15656},
  journal = {arXiv},
  note = {arXiv preprint arXiv:2506.15656}
}

@inproceedings{123,
  author = {Hao Huang and Tapan Shah and John Karigiannis and Scott Evans},
  title = {Physics and Data Collaborative Root Cause Analysis: Integrating Pretrained Large Language Models and Data-Driven {AI} for Trustworthy Asset Health Management},
  year = {2024},
  booktitle = {Annual Conference of the PHM Society},
  number = {1},
  series = {2024}
}

@article{124,
  author = {Xinyi Chen and Angelica Chen and Dean Foster and Elad Hazan},
  title = {Playing large games with oracles and ai debate},
  year = {2023},
  archivePrefix = {arXiv},
  eprint = {2312.04792},
  journal = {arXiv}
}

@inproceedings{125,
author = {Ashkinaze, Joshua and Fry, Emily and Edara, Narendra and Gilbert, Eric and Budak, Ceren},
title = {Plurals: A System for Guiding {LLM}s via Simulated Social Ensembles},
year = {2025},
isbn = {9798400713941},
publisher = {ACM},
url = {https://doi.org/10.1145/3706598.3713675},
doi = {10.1145/3706598.3713675},
booktitle = {Procs. of the 2025 CHI Conference on Human Factors in Computing Systems},
articleno = {245},
numpages = {21},
}

@inproceedings{126,
  author = {Someen Park and Jaehoon Kim and Seungwan Jin and Sohyun Park and Kyungsik Han},
  title = {{PREDICT}: multi-agent-based debate simulation for generalized hate speech detection},
  year = {2024},
  booktitle = {Procs. of the 2024 Conference on Empirical Methods in Natural Language Processing},
  pages = {20963-20987}
}

@inproceedings{127,
    title = "Preventing Rogue Agents Improves Multi-Agent Collaboration",
    author = "Barbi, Ohav  and
      Yoran, Ori  and
      Geva, Mor",
    booktitle = "Procs. of the 1st Workshop for Research on Agent Language Models",
    year = "2025",
    publisher = "ACL",
    url = "https://aclanthology.org/2025.realm-1.34/",
    pages = "486--511",
    ISBN = "979-8-89176-264-0"
}

@inproceedings{128,
  author = {Ciaran Regan and Alexandre Gournail and Mizuki Oka},
  title = {Problem-solving in language model networks},
  year = {2024},
  booktitle = {Artificial Life Conference Proceedings 36},
  number = {1},
  series = {2024},
  pages = {70}
}

@article{131,
  author = {Ali Asad and Stephen Obadinma and Radin Shayanfar and Xiaodan Zhu},
  title = {RedDebate: Safer Responses through Multi-Agent Red Teaming Debates},
  year = {2025},
  archivePrefix = {arXiv},
  eprint = {2506.11083},
  journal = {arXiv},
  note = {arXiv preprint arXiv:2506.11083}
}

@article{133,
  author = {Ruoxi Cheng and Haoxuan Ma and Shuirong Cao and others},
  title = {Reinforcement learning from multi-role debates as feedback for bias mitigation in llms},
  year = {2024},
  archivePrefix = {arXiv},
  eprint = {2404.10160},
  journal = {arXiv},
  note = {arXiv preprint arXiv:2404.10160}
}

@inproceedings{136,
    title = "Removal of Hallucination on Hallucination: Debate-Augmented {RAG}",
    author = "Hu, Wentao  and
      Zhang, Wengyu  and
      Jiang, Yiyang  and
     others",
    booktitle = "Procs. of the 63rd Annual Meeting of the Association for Computational Linguistics (Volume 1: Long Papers)",
    year = "2025",
    publisher = "ACL",
    url = "https://aclanthology.org/2025.acl-long.770/",
    doi = "10.18653/v1/2025.acl-long.770",
    ISBN = "979-8-89176-251-0"
}

@inproceedings{138,
    title = "S$^2$-{MAD}: Breaking the Token Barrier to Enhance Multi-Agent Debate Efficiency",
    author = "Zeng, Yuting  and
      Huang, Weizhe  and
      Jiang, Lei  and
      others",
    booktitle = "Procs. of the 2025 Conference of the Nations of the Americas Chapter of the Association for Computational Linguistics: Human Language Technologies (Volume 1: Long Papers)",
    year = "2025",
    publisher = "ACL",
    url = "https://aclanthology.org/2025.naacl-long.475/",
    doi = "10.18653/v1/2025.naacl-long.475",
    pages = "9393--9408",
    ISBN = "979-8-89176-189-6"
}

@inproceedings{145,
  author = {Dawei Li and Zhen Tan and Peijia Qian and others},
  title = {{SMoA}: Improving Multi-agent Large Language Models with Sparse Mixture-of-Agents},
  year = {2025},
  booktitle = {Procs. of Pacific-Asia Conference on Knowledge Discovery and Data Mining},
  pages = {54-65},
  publisher = {Springer}
}

@inproceedings{146,
  author = {Masaki Ishizaka and Akihito Taya and Yoshito Tobe},
  title = {Sparkit: A mind map-based {MAS} for idea generation support},
  year = {2024},
  booktitle = {Procs. of the International Workshop on Engineering Multi-Agent Systems},
  publisher = {Springer}
}

@article{149,
  author = {Hangfan Zhang and Zhiyao Cui and Jianhao Chen and Xinrun Wang and Qiaosheng Zhang and Zhen Wang and Dinghao Wu and Shuyue Hu},
  title = {Stop Overvaluing Multi-Agent Debate--We Must Rethink Evaluation and Embrace Model Heterogeneity},
  year = {2025},
  archivePrefix = {arXiv},
  eprint = {2502.08788},
  journal = {arXiv}
}

@inproceedings{151,
  author = {Ziqun Bao and Yu Ji and Wen Wu and Xi Chen and Liang He},
  title = {Supervisor Alignment Framework: Enhancing {LLM} Alignment with Query-Ignoring Strategy and Multi-Agent Interaction},
  year = {2025},
  booktitle = {Procs. of the 2025 IEEE International Conference on Acoustics, Speech and Signal Processing},
}

@inproceedings{154,
    title = "The Fellowship of the {LLM}s: Multi-Model Workflows for Synthetic Preference Optimization Dataset Generation",
    author = "Arif, Samee  and
      Farid, Sualeha  and
      Azeemi, Abdul Hameed  and
      Athar, Awais  and
      Raza, Agha Ali",
    booktitle = "Procs. of the Fourth Workshop on Generation, Evaluation and Metrics",
    year = "2025",
    publisher = "ACL",
    url = "https://aclanthology.org/2025.gem-1.4/",
    pages = "30--45",
    ISBN = "979-8-89176-261-9"
}

@inproceedings{160,
  author = {Xiaoxi Sun and Jinpeng Li and Yan Zhong and Dongyan Zhao and Rui Yan},
  title = {Towards detecting llms hallucination via markov chain-based multi-agent debate framework},
  year = {2025},
  booktitle = {Procs. of 2025 IEEE International Conference on Acoustics, Speech and Signal Processing},
}

@inproceedings{161,
    title = "Towards Multi-dimensional Evaluation of {LLM} Summarization across Domains and Languages",
    author = "Min, Hyangsuk  and
      Lee, Yuho  and
      Ban, Minjeong  and
      others",
    booktitle = "Procs. of the 63rd Annual Meeting of the Association for Computational Linguistics (Volume 1: Long Papers)",
    year = "2025",
    publisher = "ACL",
    url = "https://aclanthology.org/2025.acl-long.702/",
    doi = "10.18653/v1/2025.acl-long.702",
    ISBN = "979-8-89176-251-0"
}

@article{163,
  author = {Vivaan Sandwar and Bhav Jain and Rishan Thangaraj and Ishaan Garg and Michael Lam and Kevin Zhu},
  title = {Town Hall Debate Prompting: Enhancing Logical Reasoning in LLMs through Multi-Persona Interaction},
  year = {2025},
  archivePrefix = {arXiv},
  eprint = {2502.15725},
  journal = {arXiv},
  note = {arXiv preprint arXiv:2502.15725}
}

@inproceedings{164,
    title = "Tree-of-Debate: Multi-Persona Debate Trees Elicit Critical Thinking for Scientific Comparative Analysis",
    author = "Kargupta, Priyanka  and
      Agarwal, Ishika  and
      August, Tal  and
      Han, Jiawei",
    booktitle = "Procs. of the 63rd Annual Meeting of the Association for Computational Linguistics (Volume 1: Long Papers)",
    year = "2025",
    publisher = "ACL",
    url = "https://aclanthology.org/2025.acl-long.1422/",
    doi = "10.18653/v1/2025.acl-long.1422",
    pages = "29378--29403",
    ISBN = "979-8-89176-251-0"
}

@article{167,
  author = {HaoYang Shang and Xuan Liu and Zi Liang and Jie Zhang and Haibo Hu and Song Guo},
  title = {United Minds or Isolated Agents? Exploring Coordination of LLMs under Cognitive Load Theory},
  year = {2025},
  archivePrefix = {arXiv},
  eprint = {2506.06843},
  journal = {arXiv},
  note = {arXiv preprint arXiv:2506.06843}
}

@article{hendrycks2021-MMLU,
      title={Measuring Massive Multitask Language Understanding}, 
      author={Dan Hendrycks and Collin Burns and Steven Basart and others},
      year={2021},
  archivePrefix = {arXiv},
  eprint = {2009.03300},
  journal = {arXiv}
}

@inproceedings{clark-etal-2019-boolq,
    title = "{B}ool{Q}: Exploring the Surprising Difficulty of Natural Yes/No Questions",
    author = "Clark, Christopher  and
      Lee, Kenton  and
      Chang, Ming-Wei  and
      others",
    booktitle = "Procs. of the 2019 Conference of the North {A}merican Chapter of the Association for Computational Linguistics: Human Language Technologies, Volume 1 (Long and Short Papers)",
    year = "2019",
    publisher = "ACL",
    url = "https://aclanthology.org/N19-1300/",
    doi = "10.18653/v1/N19-1300"
}

@article{suzgun2022-BBH,
      title={Challenging BIG-Bench Tasks and Whether Chain-of-Thought Can Solve Them}, 
      author={Mirac Suzgun and Nathan Scales and Nathanael Schärli and others},
      year={2022},
  archivePrefix = {arXiv},
  eprint = {2210.09261},
  journal = {arXiv},
  note = {arXiv preprint arXiv:2210.09261}
}

@inproceedings{yang-etal-2018-hotpotqa,
    title = "{H}otpot{QA}: A Dataset for Diverse, Explainable Multi-hop Question Answering",
    author = "Yang, Zhilin  and
      Qi, Peng  and
      Zhang, Saizheng  and
      others",
    booktitle = "Procs. of the 2018 Conference on Empirical Methods in Natural Language Processing",
    year = "2018",
    publisher = "ACL",
    url = "https://aclanthology.org/D18-1259/",
    doi = "10.18653/v1/D18-1259",
    pages = "2369--2380"
}

@article{cobbe2021trainingverifierssolvemath,
      title={Training Verifiers to Solve Math Word Problems}, 
      author={Karl Cobbe and Vineet Kosaraju and Mohammad Bavarian and others},
      year={2021},
  archivePrefix = {arXiv},
  eprint = {2110.14168},
  journal = {arXiv},
  note = {arXiv preprint arXiv:2110.14168}
}

@article{hendrycks2021measuringmathematicalproblemsolving,
      title={Measuring Mathematical Problem Solving With the {MATH} Dataset}, 
      author={Dan Hendrycks and Collin Burns and Saurav Kadavath and others},
      year={2021},
  archivePrefix = {arXiv},
  eprint = {2103.03874},
  journal = {arXiv},
  note = {arXiv preprint arXiv:2103.03874}
}

@inproceedings{sinha-etal-2019-clutrr,
    title = "{CLUTRR}: A Diagnostic Benchmark for Inductive Reasoning from Text",
    author = "Sinha, Koustuv  and
      Sodhani, Shagun  and
      Dong, Jin  and
      Pineau, Joelle  and
      Hamilton, William L.",
    booktitle = "Procs. of the 2019 Conference on Empirical Methods in Natural Language Processing and the 9th International Joint Conference on Natural Language Processing",
    year = "2019",
    publisher = "ACL",
    url = "https://aclanthology.org/D19-1458/",
    doi = "10.18653/v1/D19-1458",
    pages = "4506--4515"
}

@misc{welbl2017crowdsourcingmultiplechoicescience,
      title={Crowdsourcing Multiple Choice Science Questions}, 
      author={Johannes Welbl and Nelson F. Liu and Matt Gardner},
      year={2017},
  archivePrefix = {arXiv},
  eprint = {1707.06209},
  journal = {arXiv}
}

@article{clark2018thinksolvedquestionanswering,
      title={Think you have Solved Question Answering? Try {ARC}, the {AI2} Reasoning Challenge}, 
      author={Peter Clark and Isaac Cowhey and Oren Etzioni and others},
      year={2018},
  archivePrefix = {arXiv},
  eprint = {1803.05457},
  journal = {arXiv},
  note = {arXiv preprint arXiv:1803.05457}
}

@inproceedings{Gururangan2020,
    title = "Don{'}t Stop Pretraining: Adapt Language Models to Domains and Tasks",
    author = "Gururangan, Suchin  and
      Marasovi{\'c}, Ana  and
      Swayamdipta, Swabha  and
      others",
    booktitle = "Procs. of the 58th Annual Meeting of the Association for Computational Linguistics",
    year = "2020",
    publisher = "ACL",
    url = "https://aclanthology.org/2020.acl-main.740/",
    doi = "10.18653/v1/2020.acl-main.740",
    pages = "8342--8360"
}

@article{Zhao2026,
  author    = {Penghao Zhao and Hailin Zhang and Qinhan Yu and Zhengren Wang and Yunteng Geng and Fangcheng Fu and Ling Yang and Wentao Zhang and Jie Jiang and Bin Cui},
  title     = {Retrieval-Augmented Generation for AI-Generated Content: A Survey},
  journal   = {Data Science and Engineering},
  year      = {2026},
  volume    = {11},
  number    = {1},
  pages      = {1--29},
  doi       = {10.1007/s41019-025-00335-5},
  url       = {https://doi.org/10.1007/s41019-025-00335-5},
  issn      = {2364-1541}
}

@article{Retzlaff2024,
author = {Retzlaff, Carl Orge and Das, Srijita and Wayllace, Christabel and Mousavi, Payam and Afshari, Mohammad and Yang, Tianpei and Saranti, Anna and Angerschmid, Alessa and Taylor, Matthew E. and Holzinger, Andreas},
title = {Human-in-the-Loop Reinforcement Learning: A Survey and Position on Requirements, Challenges, and Opportunities},
year = {2024},
publisher = {AI Access Foundation},
volume = {79},
issn = {1076-9757},
url = {https://doi.org/10.1613/jair.1.15348},
doi = {10.1613/jair.1.15348},
journal = {J. Artif. Int. Res.},
numpages = {57}
}

@INPROCEEDINGS{Zhou2016,
  author={Zhou, Xin and Jin, Yuqin and Zhang, He and Li, Shanshan and Huang, Xin},
  booktitle={2016 23rd Asia-Pacific Software Engineering Conference (APSEC)}, 
  title={A Map of Threats to Validity of Systematic Literature Reviews in Software Engineering}, 
  year={2016},
  volume={},
  number={},
  pages={153-160},
  doi={10.1109/APSEC.2016.031}}

@techreport{Kitchenham2007,
  author = {Kitchenham, Barbara Ann and Charters, Stuart},
  institution = {Keele University and Durham University Joint Report},
  number = {EBSE 2007-001},
  school = {Keele University},
  title = {Guidelines for performing Systematic Literature Reviews in Software Engineering},
  year = 2007
}

@misc{Motger2026MADReplication,
  author       = {Quim Motger and Marc Oriol and Jordi Marco and Xavier Franch},
  title        = {{nlp4se/MAD-rep-package: Replication package for "Multi-Agent Debate Strategies: Survey, Taxonomy, and Challenges"}},
  year         = {2026},
  version      = {v1.0},
  publisher    = {Zenodo},
  doi          = {10.5281/zenodo.21411493},
  url          = {https://doi.org/10.5281/zenodo.21411493}
}

\end{document}